\documentclass[11pt,a4paper]{article}

\usepackage{amsmath,amssymb,tikz,hyperref,empheq}
\usepackage{graphicx}
 \usepackage{verbatim}

\usepackage{tikz}
\usetikzlibrary {decorations.pathmorphing}
\usetikzlibrary {arrows.meta}
\usetikzlibrary{decorations.text}
 \allowdisplaybreaks

\def\be{\begin{equation}}
\def\ee{\end{equation}}
\def\ba#1\ea{\begin{align}#1\end{align}}
\def\bg#1\eg{\begin{gather}#1\end{gather}}
\def\bm#1\em{\begin{multline}#1\end{multline}}
\def\bmd#1\emd{\begin{multlined}#1\end{multlined}}

\def\a{\alpha}

\def\e{\epsilon}

\def\({\left(}
\def\){\right)}
\def\[{\left[}
\def\]{\right]}

\def \be {\begin{equation}}
\def \ee {\end{equation}}
\def \ba {\begin{array}}
\def \ea {\end{array}}
\def \bea{\begin{eqnarray}}
\def \eea{\end{eqnarray}}

\def \a {\alpha}

\def \e {\epsilon}

\def\bea{\begin{eqnarray}}
\def\eea{\end{eqnarray}}

\newcommand{\bit}{\begin{itemize}}  \newcommand{\eit}{\end{itemize}}
\newcommand{\ben}{\begin{enumerate}}  \newcommand{\een}{\end{enumerate}}

\long\def\symbolfootnote[#1]#2{\begingroup%
\def\thefootnote{\fnsymbol{footnote}}\footnote[#1]{#2}\endgroup}

\newcommand{\sysu}{{\it School of Physics and Astronomy, Sun Yat-Sen University, 2 Daxue Road, Zhuhai 519082, China}}

\begin{document}
\thispagestyle{empty}
\begin{center}

~\vspace{20pt}

{\Large\bf Island on codimension-two branes in AdS/dCFT}

\vspace{25pt}

Peng-Ju Hu, Dongqi Li and Rong-Xin Miao ${}$\symbolfootnote[1]{Email:~\sf
  miaorx@mail.sysu.edu.cn}

\vspace{10pt}${}$\sysu

\vspace{2cm}

\begin{abstract}
The previous studies of the island and double holography mainly focus on codimension-one branes. This paper explores the island on the codimension-two brane in AdS/dCFT.  The codimension-two brane is closely related to conical singularity, which is very different from the codimension-one brane. We analyze the mass spectrum of gravitons on the codimension-two brane and find that the larger the brane tension is, the smaller the gravitational mass is. The massless mode is forbidden by either the boundary or normalization conditions. We prove that the first massive gravitational mode is located on the codimension-two brane; the larger the tension, the better the localization. It is similar to the case of codimension-one brane and builds an excellent physical foundation for the study of black hole evolution on codimension-two branes. We find that the Page curve of eternal black holes can be recovered due to the island ending on the codimension-two brane. 
The new feature is that the extremal surface passing the horizon cannot be defined after some finite time in the no-island phase. Fortunately, this unusual situation does not affect the Page curve since it happens after Page time. 
\end{abstract}

\end{center}

\newpage
\setcounter{footnote}{0}
\setcounter{page}{1}

\tableofcontents

\section{Introduction}
It is widely believed that the study of quantum aspects of black holes sheds light on a consistent theory of quantum gravity. Recently, there has been a significant breakthrough toward resolving the black hole information paradox \cite{Penington:2019npb,Almheiri:2019psf}, where double holography and island play an essential role. See
\cite{Almheiri:2020cfm,Almheiri:2019hni,Rozali:2019day, Chen:2019uhq,Almheiri:2019yqk,Almheiri:2019psy,Kusuki:2019hcg,
   Balasubramanian:2020hfs,Geng:2020qvw,Chen:2020uac,Ling:2020laa,
   Kawabata:2021hac,Bhattacharya:2021jrn,Kawabata:2021vyo,
Geng:2021hlu,Krishnan:2020fer,Chen:2020hmv,Ghosh:2021axl,Omiya:2021olc,Bhattacharya:2021nqj,Geng:2021mic,Sun:2021dfl,Chou:2021boq,Ahn:2021chg,He:2021mst,Alishahiha:2020qza,Lin:2022aqf,Gan:2022jay,Omidi:2021opl,Hu:2022ymx,Azarnia:2021uch,Anous:2022wqh,Saha:2021ohr,Yadav:2022mnv} for some recent works.  Double holography is closely related to brane world theory  \cite{Randall:1999ee,Randall:1999vf,Karch:2000ct} and AdS/BCFT \cite{Takayanagi:2011zk,Fujita:2011fp,Nozaki:2012qd,Miao:2018qkc,Miao:2017gyt,Chu:2017aab,Chu:2021mvq}. Recently, a novel doubly holographic model called wedge holography has been proposed \cite{Akal:2020wfl}.  Generalizing wedge holography to codim-m defects, \cite{Miao:2021ual} proposes the so-called cone holography. Remarkably, there is a massless gravitational mode on the brane in wedge/cone holography, which is quite different from the brane world theory and AdS/BCFT. Besides, the effective theory on the brane is ghost-free higher derivative gravity and behaves like Einstein gravity in many aspects \cite{Hu:2022lxl}. See also \cite{Miao:2020oey,Geng:2020fxl,Geng:2022slq,Geng:2022tfc,Ogawa:2022fhy,Izumi:2022opi,Suzuki:2022xwv,Numasawa:2022cni,Kusuki:2021gpt} for some recent works on wedge/cone holography and AdS/BCFT. 

Previous studies of the island and double holography mainly focus on codimension-one (codim-1) branes 
\footnote{See \cite{Aghababaie:2003ar,Cline:2003ak,Hayashi:2009bt,Corradini:2001qv,deRham:2005jan} for some discussions of codimension-two branes in the framework of the brane world theory.}. In this paper, we investigate the island on codimension-two (codim-2) branes in AdS/dCFT \cite{Jensen:2013lxa,DeWolfe:2001pq}. The codim-2 brane is different from the codim-1 brane in many aspects. First, as a minimal surface, the codim-2 brane is always perpendicular to the AdS boundary, while this is not the case for codim-1 brane.  Second, the codim-2 brane is closely related to conical singularity, which has critical applications in holographic entanglement entropy \cite{Ryu:2006bv} and holographic R\'enyi entropy \cite{Dong:2016fnf}. Third, the tension of codim-2 brane has to be a constant in Einstein gravity, while the tension of codim-1 brane can be a function. To allow more general tension and matter fields on codim-2 brane, one can consider Gauss-Bonnet gravity \cite{Bostock:2003cv}. Fourth, there is no well-defined thin-brane limit in Einstein gravity for codimensions higher than two \cite{Bostock:2003cv}. Thus we focus on codim-2 branes in this paper and leave the study of higher codimensional branes in future work. 

Since codim-2 branes are very different from codim-1 branes, it is interesting to explore the island mechanism to see if there are new features on codim-2 branes. Recent work finds that the localization of massless gravity is impossible for codim-2 branes with at least one non-compact extra dimension \cite{Li:2020mgc}. This seems to rule out the possibility of studying the black hole information paradox and island on codim-2 branes. One of the motivations of this paper is to show this is not the case. In fact, similar to the codim-1 brane \footnote{This paper focuses on the Karch-Randall brane, where the induced geometry on the brane is an AdS space.}, we find that there is no massless gravity on the codim-2 brane. Instead, the gravity is massive and can indeed be located on the codim-2 brane. 

Let us summarize our main results below. 
In this paper, we investigate the mass spectrum of gravitons and island on the codim-2 brane in AdS/dCFT. We find that the mass spectrum is positive and there is no massless mode, which is similar to the case of AdS/BCFT. Interestingly, the mass spectrum becomes continuous in the large tension limit, which is different from AdS/BCFT. Furthermore, we prove that the first massive gravitational mode is located on the codim-2 brane; the larger the tension, the better the localization. It is similar to the case of AdS/BCFT and builds a solid physical foundation for studying the island on codim-2 branes. 
Finally, we carefully investigate eternal hyperbolic black holes and AdS black holes on the codim-2 brane and find that the Page curves of these black holes can be recovered due to the island ending on the codim-2 brane. Remarkably, the extremal surface passing the horizon cannot be defined after some finite time in the no-island phase. However, since this unusual situation happens only after Page time, it does not affect the time evolution of entanglement entropy. 

The paper is organized as follows. 
In section 2, we review AdS/dCFT and double holography with codim-2 defects. Section 3 discusses the mass spectrum of gravitons on the codim-2 brane and verifies that the massive gravity can be located on the brane. In section 4, we study a toy model in AdS$_4$/dCFT$_3$ and find that the Page curve of the eternal black hole can be recovered due to the island ending on the codim-2 brane. Compared with the cases in higher dimensions, this toy model can obtain more analytical results. In this toy model, we clearly show why the extremal surface passing through the horizon cannot be defined after some finite time in the no-island phase and why this does not affect the Page curve. In section 5, we generalize the discussions to AdS/dCFT in higher dimensions and find that the qualitative behavior of the Page curve is the same as that of the toy model. Finally, we conclude with some open problems in section 6.

\section{Review of AdS/dCFT}

\begin{figure}[t]
\centering
\includegraphics[width=6.1cm]{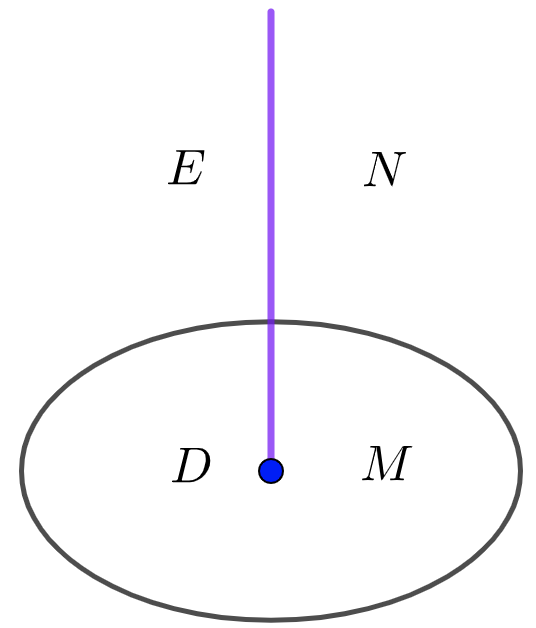}
\caption{Geometry of AdS/dCFT. $N$ is the 
AdS$_{d+1}$ space in the bulk, $M=\partial N$ is the AdS boundary, $E$ (purple line) is the codim-2 brane in the bulk, $D=\partial E$ (blue point) is the codim-2 defect on the AdS boundary. AdS/dCFT conjectures that the gravity coupled with a brane $E$ in the bulk $N$ is dual to the CFT coupled with a defect $D$ on the AdS boundary $M$. }
\label{AdSdCFT}
\end{figure}

In this section, we give a brief review of AdS/dCFT with codim-2 defects. See Fig. \ref{AdSdCFT} for the geometry, where $N$ is the AdS$_{d+1}$ space in the bulk, $M=\partial N$ is the AdS boundary, $E$ (purple line) is the codim-2 brane in the bulk and $D=\partial E$ (blue point) is the codim-2 defect on the AdS boundary. AdS/dCFT conjectures that the gravity coupled with a brane $E$ in bulk $N$ is dual to the CFT coupled with a defect $D$ on the AdS boundary $M$.
In the framework of brane world or double holography, 
our world with dynamical gravity (black hole) is defined on the brane $E$, and the CFT without gravity (bath) lives on the AdS boundary $M$. 

The action of AdS/dCFT is given by
\begin{eqnarray}\label{action}
  I=\frac{1}{16\pi G_N}\int_N  dX^{d+1}\sqrt{|G|} (R-2\Lambda)
-T_E \int_E  dy^{d-1}\sqrt{|h|},
\end{eqnarray}
where $G_N$ is Newton's constant, $R$ is the Ricci scalar, $-2\Lambda=d(d-1)$ is the cosmological constant (we have set AdS radius $L=1$), $T_E$ is the tension of codim-2 brane $E$, 
$X^A=(x^1,x^2, y^i)$ and $y^i$ are coordinates in bulk $N$ and on the brane $E$, respectively. Similarly, $G_{AB}$ and $h_{ij}$ are metrics on $N$ and $E$, respectively. Taking variations of the action (\ref{action}), we get equations of motion (EOM)
\begin{eqnarray}\label{gravityEOM}
R_{AB}-\frac{R-2\Lambda}{2} G_{AB}=-8\pi G_N T_E \ 
\frac{\sqrt{|h|}}{\sqrt{|G|}} h^i_A h^j_B h_{ij} \ \delta^{(2)}(x^1,x^2),
\end{eqnarray}
where $h^i_A=\partial y^i/\partial x^A$ is the projection operator and $(x^1, x^2)$ denote the directions normal to the brane $E$.

Let us take the tensionless brane as an example to illustrate the geometry of AdS/dCFT.  The bulk AdS metric is given by 
\begin{eqnarray}\label{AdSdCFTmetric}
ds^2=dr^2+\sinh^2(r) d\theta^2+\cosh^2(r) \frac{dz^2-dt^2+\sum_{a=1}^{d-3}dy_a^2}{z^2},
\end{eqnarray}
where we have set the AdS radius $L=1$, $r$ denotes the proper distance to the brane, $r=0$ and $r=\infty$ denotes the locations of brane $E$ and the AdS boundary $M$, respectively. The codim-2 defect $D$ is located at $z=0$. There is no conical singularity on the defect $D$ for the tensionless brane. This is natural since the tensionless brane has no back-reaction to the AdS space.  Let us explain more on this point. The bulk spacetime should be smooth in order to satisfy Einstein equations. This means that there is no conical singularity near the brane $E$ located at $r=0$. From (\ref{AdSdCFTmetric}) with $r\sim 0$, we determine the period of $\theta$ to be $2\pi$.  On the other hand, the induced metric on the AdS boundary $r\to \infty$ is conformally equivalent to
\begin{eqnarray}\label{AdSMmetric}
ds_M^2\sim z^2 d\theta^2+dz^2-dt^2+\sum_{a=1}^{d-3}dy_a^2. 
\end{eqnarray}
Clearly, the induced metric has no conical singularity for $\theta$ with period $2\pi$.

Let us go on to discuss the brane with non-zero tension. The corresponding metric is given by \cite{Miao:2021ual}
\begin{eqnarray}\label{metricwithtension}
ds^2=dr^2+f(r) d\theta^2+g(r) \frac{dz^2-dt^2+\sum_{a=1}^{d-3}dy_a^2}{z^2},
\end{eqnarray}
where $f(r)$ and $g(r)$ obey EOMs  (\ref{fgFrelations1},\ref{fgFrelations2}) and the boundary conditions (BCs)
\begin{eqnarray}\label{fgBC1}
&&\lim_{r\to 0} f(r)= \frac{r^2}{q^2}, \ \  \lim_{r\to 0} g(r) \ \text{ is \ finite},\\
&& \lim_{r\to \infty} f(r)= \lim_{r\to \infty} g(r)\to \infty. \label{fgBC2}
\end{eqnarray}
Here $q\ge 1$ is a positive constant. In general, (\ref{metricwithtension}) is no longer an AdS space due to the back-reaction of the brane. And this leads to a conical singularity generally. 
Near the brane $E$, we have 
\begin{eqnarray}\label{tensionmetricE}
\lim_{r\to 0} ds^2=dr^2+ \frac{r^2}{q^2} d\theta^2+...
\end{eqnarray}
Regularity at $r=0$ fixes the angle period to be
\begin{eqnarray}\label{tensionperiod}
\theta \simeq \theta+2\pi q. 
\end{eqnarray}
From (\ref{metricwithtension})(\ref{fgBC2}), we notice that the induced metric on the AdS boundary is still conformally equivalent to (\ref{AdSMmetric}). As a result, there is a conical singularity on the defect $D$ with the angle period (\ref{tensionperiod}) unless $q=1$. Recall that the brane tension is given by 
\begin{eqnarray}\label{tension}
8\pi G_N T_E= 2\pi (1-\frac{1}{q}).
\end{eqnarray}
Thus a brane with non-zero tension causes a conical singularity on the defect.

Solving Einstein equations together with BCs (\ref{fgBC1},\ref{fgBC2}), we obtain for $d=4$
\begin{eqnarray}\label{f4d}
&&f(r)= \frac{\left(1-2\bar{r}_h^2\right)^2 \sinh ^2(2 r)}{\left(4\bar{r}_h^2-2\right) \cosh (2 r)+2}, \\ \label{g4d}
&&g(r)=\frac{1}{2} \left(\left(2 \bar{r}_h^2-1\right) \cosh (2 r)+1\right),
\end{eqnarray}
where $\bar{r}_h$ is a constant will be defined below. 
As for general $d$, there is no analytical solutions to $f(r)$ and $g(r)$ for the brane with non-zero tension. Performing the radial coordinate transformation
\begin{eqnarray}\label{rbarr}
dr=\frac{d\bar{r}}{\sqrt{F(\bar{r})}},
\end{eqnarray}
we get an analytical bulk metric in general dimensions
\begin{eqnarray}\label{rbarmetric}
ds^2=\frac{d\bar{r}^2}{F(\bar{r})}+ F(\bar{r}) d\theta^2+\bar{r}^2 \frac{dz^2-dt^2+\sum_{a=1}^{d-3}dy_a^2}{z^2},
\end{eqnarray}
where 
\begin{eqnarray}\label{rbarF}
F(\bar{r})=\bar{r}^2-1-\frac{\bar{r}_h^{d-2}(\bar{r}_h^{2}-1)}{\bar{r}^{d-2}},
\end{eqnarray}
with 
\begin{eqnarray}\label{rbarhorizon}
\bar{r}_h=\frac{1+\sqrt{d^2 q^2-2 d q^2+1}}{d q}.
\end{eqnarray}
Note that the brane $E$ is located at $\bar{r}=\bar{r}_h$. 

Comparing (\ref{metricwithtension}) with (\ref{rbarmetric}), we read off
\begin{eqnarray}\label{fgFrelations}
f(r)=F(\bar{r}) , \ \ g(r)=\bar{r}^2,
\end{eqnarray}
which together with (\ref{rbarF}) yields 
\begin{eqnarray}\label{fgFrelations1}
f(r)=g(r)-1-\frac{\bar{r}_h^{d-2}(\bar{r}_h^{2}-1)}{g(r)^{(d-2)/2}}.
\end{eqnarray}
The $\theta\theta$ component of Einstein equations gives
\begin{eqnarray}\label{fgFrelations2}
-4 g(r) \left(d+g''(r)-2\right)-(d-4) g'(r)^2+4 d g(r)^2=0.
\end{eqnarray}
Now we obtain the EOMs for $f(r)$ and $g(r)$.

To end this section, let us make some comments. First, the codim-2 defect $D$ is related to the conical singularity, which is different from the codim-1 defect, such as the interface and boundary. Second, AdS/dCFT with a codim-2 defect is closely related to holographic R\'enyi entropy. Third, the bulk spacetime (\ref{metricwithtension},\ref{rbarmetric}) take the form 
$\text{C}_2\times \text{AdS}_{d-1}$, where $\text{C}_2$ denotes 
a two-dimensional cone. The $\text{AdS}_{d-1}$ subspace of  (\ref{metricwithtension},\ref{rbarmetric}) can be replaced by general negatively curved Einstein manifolds such as 
hyperbolic black holes and AdS black holes \cite{Miao:2021ual}.  For example, the bulk metric  (\ref{metricwithtension}) can be generalized to
\begin{eqnarray}\label{BHmetricr}
ds^2=dr^2+f(r) d\theta^2+g(r) ds^2_{\text{BH}},
\end{eqnarray}
or equivalently
\begin{eqnarray}\label{BHmetricrbar}
ds^2=\frac{d\bar{r}^2}{F(\bar{r})}+ F(\bar{r}) d\theta^2+\bar{r}^2 ds^2_{\text{BH}},
\end{eqnarray}
with
\begin{eqnarray}\label{BH on brane}
ds^2_{\text{BH}}=\begin{cases}
\frac{1 }{z^2}\Big( \frac{dz^2}{1-z^{2}}-(1-z^{2})dt^2+ dH_{d-3}^2\Big),
\  \ \ \ \text{hyperbolic black hole}, \\
\frac{1 }{z^2}\Big( \frac{dz^2}{1-z^{d-2}}-(1-z^{d-2})dt^2+\sum_{a=1}^{d-3}dy_a^2\Big),
 \ \ \ \text{AdS black hole},
\end{cases}
\end{eqnarray}
where $f(r), g(r)$ obeys EOMs (\ref{fgFrelations1},\ref{fgFrelations2}), $F(\bar{r})$ is given by (\ref{rbarF}), $dH_{d-3}^2$ denotes the line element on a $(d-3)$-dimensional hyperbolic plane with unit curvature and the black hole horizon is at $z=1$. Fourth, (\ref{BHmetricr},\ref{BHmetricrbar}) are the typical metrics we used in the study of Page curve, where the black hole is located on the codim-2 brane and the bath is on the AdS boundary.

\section{Massive gravity on codim-2 brane}

In this section, we analyze the gravitons' mass spectrum on the codim-2 brane in AdS/dCFT. Similar to the case of AdS/BCFT, we find that the mass spectrum is positive, and there is no massless gravity on the brane. In the large tension limit, the first massive gravitational mode is almost massless and is well-located on the codim-2 brane. Thus there are solid physical foundations for studying the black hole evolution and island on codim-2 branes. 

\subsection{Mass spectrum}

We take the following ansatz of the perturbation metric
\begin{eqnarray}\label{perturbationmetric}
ds^2=dr^2+f(r) d\theta^2+g(r) \Big( h^{(0)}_{ij}(y)+ \epsilon \ H(r)  h^{(1)}_{ij}(y) \Big) dy^i dy^j,
\end{eqnarray}
where $h^{(0)}_{ij}$ is the AdS metric on the brane located at $r=0$ and $ \epsilon$ denotes the order of perturbation. We are interested in the gravitational modes, thus we focus on the transverse and traceless gauge
\begin{eqnarray}\label{hij1gauge}
D^i h^{(1)}_{ij}=0,\ \ \  h^{(0)ij}h^{(1)}_{ij}=0,
\end{eqnarray}
where $D^i$ is the covariant derivative with respect to
 $h^{(0)}_{ij}$.  Substituting (\ref{perturbationmetric},\ref{hij1gauge}) into Einstein equations and separating variables, we obtain
 \begin{eqnarray}\label{EOMmassivehij}
&& \left(D_i D^i+2-m^2\right) h^{(1)}_{ij}(y)=0,\\
&&2 H''(r)+H'(r) \left(\frac{(d-1) g'(r)}{g(r)}+\frac{f'(r)}{f(r)}\right)+\frac{2 m^2 }{g(r)} H(r)=0, \label{EOMmassiveH}
\end{eqnarray}
where $m$ denotes the mass of gravitons on the brane, which will be determined later.  We impose the natural boundary condition on the codim-2 brane
 \begin{eqnarray}\label{spectrum:naturalBC}
H(0) \ \ \text{is \ finite},
\end{eqnarray}
and the standard Dirichlet boundary condition (DBC) on the AdS boundary 
 \begin{eqnarray}\label{spectrum:DBC}
\text{DBC}:\ H(\infty)=0.
\end{eqnarray}
Clearly, (\ref{EOMmassivehij}) shows that there are massive gravitons on the brane. They have a natural physical origin, which 
is just the Kaluza-Klein modes.  
From EOM (\ref{EOMmassiveH}) and BCs (\ref{spectrum:naturalBC},\ref{spectrum:DBC}), one can determine the mass spectrum.  

We normalize $H(r)$ by the orthogonal condition
\begin{eqnarray}\label{3.1:normalizedH}
\int_0^{\infty} dr f(r)^{\frac{1}{2}}  g(r)^{\frac{d-3}{2}} H_m(r) H_{m'}(r)=\int_{\bar{r}_h}^{\infty} d\bar{r}  \bar{r}^{d-3} H_m( \bar{r}) H_{m'}( \bar{r})=\delta_{m,m'},
\end{eqnarray}
where $m, m'$ denote the masses of gravitons. The normalizable condition (\ref{3.1:normalizedH}) is necessary for the localization of gravity on the brane since it implies that the wave function $H(r)$ falls off quickly enough when it goes far from the brane, i.e., $r\to \infty$.

\subsubsection{Tensionless case}

 Let us first study the tensionless case with $f(r)=\sinh^2(r)$ and $g(r)=\cosh^2(r)$.  In this case, the EOM (\ref{EOMmassiveH})  becomes
 \begin{eqnarray}\label{spectrum:EOMHtensionless}
H''(r)+\tanh (r) \left(d+\text{csch}^2(r)\right) H'(r)+m^2 H(r) \text{sech}^2(r)=0,
\end{eqnarray}
which can be solved as \cite{Miao:2021ual}
 \begin{eqnarray}\label{EOMMBCmassiveHsolution}
H(r)=c_1 \, _2F_1\left(a_1,a_2;1;\tanh ^2(r)\right)+c_2 G_{2,2}^{2,0}\left(\tanh ^2(r)|
\begin{array}{c}
 a_1+\frac{d}{2}, a_2+\frac{d}{2} \\
 0,0 \\
\end{array}
\right),
\end{eqnarray}
where $_2F_1$ is the hypergeometric function, $G_{2,2}^{2,0}$ is the Meijer G function, $c_1$ and $c_2$ are integral constants and $a_i$ are given by
 \begin{eqnarray}\label{aibia1}
&&a_1=\frac{1}{4} \left(2-d-\sqrt{(d-2)^2+4 m^2}\right),\\ \label{aibia2}
&&a_2=\frac{1}{4} \left(2-d+\sqrt{(d-2)^2+4m^2}\right). 
\end{eqnarray}
Near the brane $E$ ($r=0$), (\ref{EOMMBCmassiveHsolution}) behaves as
 \begin{eqnarray}\label{EOMMBCmassiveHseries}
H(r)\sim c_2 \ln r+O(r^0),
\end{eqnarray}
From the natural BC (\ref{spectrum:naturalBC}) on brane $E$, we get
 \begin{eqnarray}\label{spectrum:c2}
c_2=0.
\end{eqnarray}
Imposing DBC (\ref{spectrum:DBC}) on AdS boundary, we get
\begin{eqnarray}\label{spectrum:DBC1}
H(\infty)&=&c_1\  {}_2F_1\left(a_1,a_2;1;1\right) \nonumber\\
&=&\frac{c_1 \Gamma \left(\frac{d}{2}\right)}{\Gamma \left(\frac{d+2}{4}-\frac{1}{4} \sqrt{(d-2)^2+4 m^2}\right) \Gamma \left(\frac{d+2}{4}+\frac{1}{4} \sqrt{(d-2)^2+4 m^2}\right)}=0.
\end{eqnarray}
From the above equation, we obtain the mass spectrum of gravitons 
\begin{eqnarray}\label{spectrum:tensionlessmm}
m^2=2 k (2 k+d-2), \ \ \ \text{for zero brane tension},
\end{eqnarray}
where $k=1,2,...$ are positive integers.  Note that the massless mode $m^2=0$ is not a solution to (\ref{spectrum:DBC1}). 
Thus the massless mode is ruled out by DBC (\ref{spectrum:DBC}) on the AdS boundary. It should be mentioned that the massless mode with $H(r)=c_1$ is a solution to the Neumann boundary condition (NBC) 
 \begin{eqnarray}\label{spectrum:NBC}
\text{NBC}:\ H'(\infty)=0.
\end{eqnarray}
However, this solution is non-normalizable
\begin{eqnarray}\label{3.1.1:normalizedH massless}
\int_0^{\infty} dr \sinh(r) \cosh^{d-3}(r)  H_0^2(r)= \int_0^{\infty} dr \sinh(r) \cosh^{d-3}(r)  c^2_1\to \infty.
\end{eqnarray}
Thus there is no way to have a massless mode located on the brane in AdS/dCFT, which is similar to AdS/BCFT.

\subsubsection{Large tension limit}

Let us go on to study the large tension limit $q\to \infty$ (\ref{tension}).  For simplicity, we focus on the case $d=4$, where the analytic solutions to $f(r)$ and $g(r)$ are known. See (\ref{f4d},\ref{g4d}).  We comment on the case of general dimensions at the end of this subsection. 

Substituting  (\ref{f4d},\ref{g4d}) into (\ref{EOMmassiveH}) with $d=4$, we get EOM 
 \begin{eqnarray}\label{spectrum:EOMH4d}
2 H''(r) \left(\left(2 \bar{r}_h^2-1\right) \cosh (2 r)+1\right)+4 \text{csch}(2 r) H'(r) \left(\left(2 \bar{r}_h^2-1\right) \cosh (4 r)+\cosh (2 r)\right)+4 m^2 H(r)=0.
\end{eqnarray}
From  (\ref{rbarhorizon}), we have $\bar{r}_h^2=1/2$ for $d=4$ and $q\to \infty$. Then (\ref{spectrum:EOMH4d}) becomes
 \begin{eqnarray}\label{spectrum:EOMHlarged}
H''(r)+2 \coth (2 r) H'(r)+2 m^2 H(r)=0,
\end{eqnarray}
in the large tension limit $q\to \infty$.  Solving the above equation, we get
 \begin{eqnarray}\label{spectrum:Hforlarged}
H(r)=c_1 P_{\lambda}\left(\cosh(2r)\right)+c_2 Q_{\lambda}\left(\cosh(2r)\right),
\end{eqnarray}
where $P_{\lambda},Q_{\lambda}$ denote the Legendre function with
 \begin{eqnarray}\label{spectrum:lambda}
\lambda=\frac{1}{2} \sqrt{1-2 m^2}-\frac{1}{2}. 
\end{eqnarray}
 Imposing the natural BC (\ref{spectrum:naturalBC}) on the brane $E \ (r=0)$, we derive $c_2=0$.  Then $H(r)$ becomes
 \begin{eqnarray}\label{spectrum:Hforlarged1}
H(r)=c_1 P_{\lambda}\left(\cosh(2r)\right).
\end{eqnarray}
Remarkably, (\ref{spectrum:Hforlarged1}) automatically obeys the DBC (\ref{spectrum:DBC}) on AdS boundary for any positive $m^2$.  This means that the mass spectrum is continuous in the large tension limit
\begin{eqnarray}\label{spectrum:mmforlargetension}
m^2>0, \ \ \ \text{is continuous for large tension with $q\to \infty$},
\end{eqnarray}
which is quite different from the case of AdS/BCFT with
 a codim-1 brane.  
Note that the massless mode with $H(r)=c_1$ does not satisfy the DBC (\ref{spectrum:DBC}) on the AdS boundary.  Note also that, although the massless mode with $H(r)=c_1$ obeys NBC (\ref{spectrum:NBC}), it is non-normalizable
\begin{eqnarray}\label{3.1.2:normalizedH massless}
\int_0^{\infty}  f(r)^{\frac{1}{2}}  g(r)^{\frac{d-3}{2}} H_0(r)^2 dr = \int_{\bar{r}_h}^{\infty}  c^2_1 \bar{r}^{d-3}  d\bar{r}\to \infty.
\end{eqnarray}
and thus is not allowed.  Thus, the lightest mass approaches zero but cannot be zero in the large tension limit.

In the above discussions, we focus on $d=4$. Let us comment on the case in general dimensions. Although the exact expressions of $f(r)$ and $g(r)$ are unknown for $d\ne 4$,  in the large tension limit $q\to \infty$, we have
\begin{eqnarray}\label{spectrum:fglargelimit}
f(r)= g'(r)\to  0, \ \ g(r) \to \frac{d-2}{d}, \ \ \frac{f'(r)}{f(r)} \to 2 \sqrt{d} \coth \left(\sqrt{d} \ r\right).
\end{eqnarray}
Substituting the above limits into (\ref{EOMmassiveH}), we get a well-defined equation
 \begin{eqnarray}\label{spectrum:EOMHlargegenerald}
H''(r)+  \sqrt{d} \coth(\sqrt{d} \ r)H'(r)+\frac{d m^2 }{d-2}H(r)=0.
\end{eqnarray}
Solving (\ref{spectrum:EOMHlargegenerald}) together with BCs (\ref{spectrum:naturalBC},\ref{spectrum:DBC}), we obtain
 \begin{eqnarray}\label{spectrum:Hforlargedgenerald}
H(r)=c_1 P_{\frac{1}{2} \left(\sqrt{1-\frac{4 m^2}{d-2}}-1\right)}\left(\cosh \left(\sqrt{d} r\right)\right),
\end{eqnarray}
with the same mass spectrum (\ref{spectrum:mmforlargetension}) as the case of $d=4$.

To summarize, the mass spectrum on the codim-2 brane is continuous and positive in the large tension limit in general dimensions. The massless mode is forbidden by either the DBC (\ref{spectrum:DBC}) or the normalizable condition (\ref{3.1:normalizedH}). 

\subsubsection{General tension}

Let us go on to discuss the mass spectrum for general tensions. Since it is difficult to find analytical solutions, we focus on numeral calculations in this subsection.  We first 
work in  the coordinate $\bar{r}$ (\ref{rbarr}), where the exact bulk metric (\ref{rbarmetric}) is known in general dimensions. In coordinate $\bar{r}$, the EOM (\ref{EOMmassiveH}) becomes
\begin{eqnarray}\label{EOMHrbar}
H''\left(\bar{r}\right)+\left(\frac{d-1}{\bar{r}}+\frac{F'\left(\bar{r}\right)}{F\left(\bar{r}\right)}\right)H'\left(\bar{r}\right) +\frac{m^2 }{\bar{r}^2 F\left(\bar{r}\right)}H\left(\bar{r}\right)=0,
\end{eqnarray}
where $F\left(\bar{r}\right)$ is given by (\ref{rbarF}),  $\bar{r}\ge \bar{r}_h$ and $\bar{r}=\bar{r}_h$ denotes the location of the brane.  We impose the natural BC on the brane, which means that 
$H(\bar{r}_h)$ is finite,
\begin{eqnarray}\label{naturlaBCrbar}
H(\bar{r})= 1+ \sum_{i=1} a_i (\bar{r}- \bar{r}_h)^i,
\end{eqnarray}
where $a_i$ are some constants to be determined. 
Substituting (\ref{naturlaBCrbar}) into (\ref{EOMHrbar}), we solve 
\begin{eqnarray}\label{naturlaBCa1}
&&a_1=\frac{-m^2}{\bar{r}_h \left(d \bar{r}_h^2-d+2\right)},\\ \label{naturlaBCa2}
&& a_2 =\frac{m^2 \left(4 d \bar{r}_h^2-2 d+m^2+4\right)}{4 \bar{r}_h^2 \left(d \left(\bar{r}_h^2-1\right)+2\right){}^2},\\
&&...\nonumber
\end{eqnarray}
where $\bar{r}_h$ (\ref{rbarhorizon}) depends on the brane tension (\ref{tension}) which can be labelled by $q$. 
For any given $m^2$, now we can numerically solve (\ref{EOMHrbar}) with the BC
\begin{eqnarray}\label{3.1.3:BConhorizon}
H(\epsilon)= 1+ a_1 \e + a_2 \e^2,\  H'(\epsilon)= a_1+ 2 a_2 \e,
\end{eqnarray}
where $\epsilon$ is a cut-off near the brane. Unless we choose the mass suitably, in general, the numerical solution does not obey the DBC on the AdS boundary  
\begin{eqnarray}\label{3.1.3:BConAdSbdy}
H(\bar{r}_c)= 0,
\end{eqnarray}
where $\bar{r}_c$ is an UV cut-off. 
Naturally, we can use the shooting method to determine the mass spectrum: we adjust the input $m^2$ so that the solution derived from (\ref{EOMHrbar}) and (\ref{3.1.3:BConhorizon}) satisfies the DBC (\ref{3.1.3:BConAdSbdy}).   

Without loss of generality, we choose $\e=0.0001$ and  $\bar{r}_c=1000$. By applying the shooting method, we derive the mass spectrum for different brane tension labeled by $q$. See Table \ref{table1spectrum} for the mass of the first mode, which decreases with the brane tension. See also Table \ref{table1spectrum2} and Table \ref{table1spectrum3} for the mass spectrum of the first ten modes, where the mass also decreases with the tension. 

\begin{table}[ht]
\caption{Mass spectrum for general tension labeled by $q$}
\begin{center}
    \begin{tabular}{| c | c | c | c |  c | c | c | c| c| c|c| }
    \hline
   $q$  & $1$ & $2$ & 3  & 4& 5 & 6& 7&8&9&10\\ \hline
  $m^2$ for $d=4$   & 8 & 4.44 & 3.40 & 2.89 & 2.58 & 2.37 &2.22&2.10&2.00&1.92 \\ \hline
  $m^2$ for $d=5$  & 10 & 5.94 & 4.68 & 4.03& 3.63 &3.36&3.15&2.99&2.87&2.76 \\ \hline
    \end{tabular}
\end{center}
\label{table1spectrum}
\end{table}
\begin{table}[ht]
	\caption{Mass spectrum of general gravitational modes labeled by $n$ for $d=4$}
	\begin{center}
		\begin{tabular}{| c | c | c | c |  c | c | c | c| c| c|c| }
			\hline
			$n$  & $1$ & $2$ & 3  & 4& 5 & 6& 7&8&9&10\\ \hline
			$m_n^2$ for $q=1$   & 8 & 24 & 48 & 80 & 120 & 168 &224&288&360&440 \\ \hline
			$m_n^2$ for $q=5$  & 2.58 & 7.56 & 15.05 & 25.04& 37.53 &52.53&70.02&90.01&112.50&137.48 \\ \hline
			$m_n^2$ for $q=10$  & 1.92 & 5.47 & 10.83 & 17.99& 26.94 &37.67&50.20&64.52&80.63&98.53 \\ \hline
		\end{tabular}
	\end{center}
	\label{table1spectrum2}
\end{table}
\begin{table}[ht]
	\caption{Mass spectrum of general gravitational modes labeled by $n$ for $d=5$}
	\begin{center}
		\begin{tabular}{| c | c | c | c |  c | c | c | c| c| c|c| }
			\hline
			$n$  & $1$ & $2$ & 3  & 4& 5 & 6& 7&8&9&10\\ \hline
			$m_n^2$ for $q=1$   & 10 & 28 & 54 & 88 & 130 & 180 &238&304&378&460 \\ \hline
			$m_n^2$ for $q=5$  & 3.63 & 10.16 & 19.66 & 32.10& 47.47 &65.77&87.01&111.17&138.26&168.28 \\ \hline
			$m_n^2$ for $q=10$  & 2.76 & 7.56 & 14.60 & 23.83& 35.24 &48.83&64.60&82.53&102.65&124.94 \\ \hline
		\end{tabular}
	\end{center}
	\label{table1spectrum3}
\end{table}

Let us go on to discuss the numerical calculation in coordinate $r$, which provides a double-check of our results. 
Note that the above numerical calculation in  coordinate $\bar{r}$ does not work well in the large tension limit $q\to \infty$. That is because $a_i$ (\ref{naturlaBCa1},\ref{naturlaBCa2}) becomes infinite in the large tension limit with $\bar{r}_h^2\to (d-2)/d$.  It behaves better in the coordinate $r$ (\ref{rbarr}), where the expansion coefficients of $H(r)$ near the brane are finite in the large tension limit.  Solving EOMs (\ref{fgFrelations1},\ref{fgFrelations2},\ref{EOMmassiveH}) near the brane $r=0$, we get
\begin{eqnarray}\label{3.1.3:nearbranefr}
&&f(r)=\frac{r^2}{q^2}+ O(r^4),\\ \label{3.1.3:nearbranegr}
&&g(r)=\bar{r}^2_h+ b_1 r^2+ b_2 r^4+O(r^6),\\ \label{3.1.3:nearbraneHr}
&&H(r)=1+ c_1 r^2+ c_2 r^4+O(r^6),
\end{eqnarray}
where 
\begin{eqnarray}\label{3.1.3:bici}
&&b_1=1+\frac{1}{2} d \left(\bar{r}_h^2-1\right), \ b_2=\frac{1}{48} \left(-(d-6) d^2 \bar{r}_h^2-\frac{(d-4) (d-2)^2}{\bar{r}_h^2}+2 d \left(d^2-7 d+10\right)\right),\nonumber\\
&&c_1=-\frac{m^2}{4 \bar{r}_h^2}, \ \ \ c_2=\frac{m^2 \left(2 (d+3) d \bar{r}_h^2-2 d^2+3 m^2+8\right)}{192 \bar{r}_h^4}.
\end{eqnarray}
Interestingly, $b_i, c_i$ are finite in the large tension limit with $\bar{r}_h^2\to (d-2)/d$. 

Following the above approach, we can solve numerically $f(r),g(r), H(r)$ and then determine the mass spectrum by using the  shooting method.  We recover exactly the results shown in 
Table. \ref{table1spectrum}, Table. \ref{table1spectrum2} and Table. \ref{table1spectrum3}. This is a double check of our results. 
Note that one should take larger and larger cut-off $r_c$ near the AdS boundary as $q$ increases in order to preserve the numerical precision. 

In summary, in this subsection, we investigate the mass spectrum of gravitons on the codim-2 brane. We find that the mass spectrum is always positive, and the massless mode is forbidden by either the DBC or the normalizable condition. Furthermore, we precisely work out the mass spectrum in the small and large tension limits. Remarkably, the mass spectrum becomes continuous in the large tension limit, which is different from the case of the codim-1 brane. For general brane tension, we analyze the mass spectrum numerically and find that the larger the tension is, the smaller the mass is. In the large tension limit, the lightest mass approaches zero but cannot be zero. This is similar to the case of the codim-1 brane.

\subsection{Localized gravity}

By analyzing the ``wave function" and the effective potential, we show that the first massive gravitational mode is located on the brane. The larger the brane tension is, the better the localization is. 

\subsubsection{Wave function}

For simplicity, we work with the coordinate $r$, where $r$ is the proper distance to the brane. By `` localization", we means that the wave function $H(r)$ defined in the perturbative metric (\ref{perturbationmetric}) peaks on the brane only and decays when it goes far from the brane. 

By applying the numerical methods of sect.3.1.3, we can solve $H(r)$ for various brane tensions and spacetime dimensions. We normalize $H(r)$ by the orthogonal condition (\ref{3.1:normalizedH}). For the tensionless case, the above normalization fixes $H(r)$ (\ref{EOMMBCmassiveHsolution},\ref{spectrum:c2}) to be
\begin{eqnarray}\label{3.2.1:normalizedHzerotension}
H(r)=\sqrt{4k+d-2}\ _2F_1\left(a_1,a_2;1;\tanh ^2(r)\right),
\end{eqnarray}
where $a_1, a_2$ are given by (\ref{aibia1},\ref{aibia2}) and the positive integer $k$ labels the mass spectrum (\ref{spectrum:tensionlessmm}). For general brane tension, one can fix $H(r)$ by numeral calculations.

\begin{figure}[t]
\centering
\includegraphics[width=10cm]{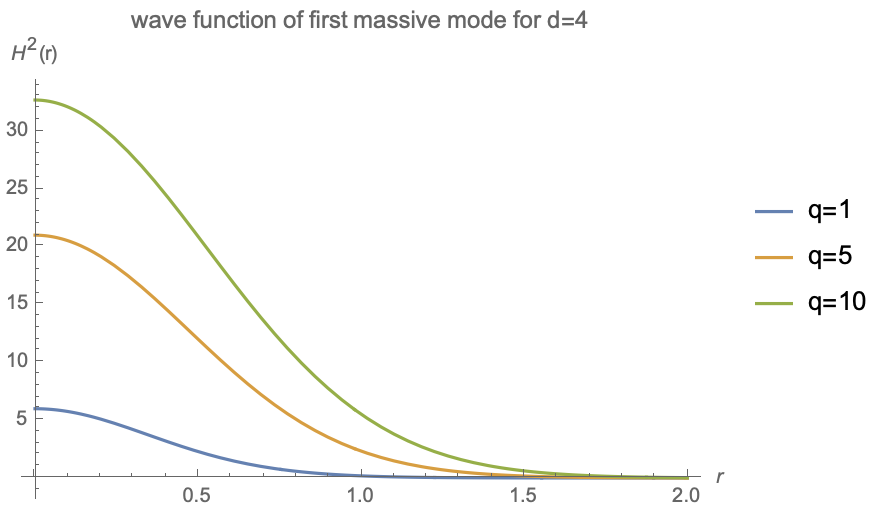}
\caption{Wave function $H^2(r)$ of the first massive gravitational mode for tensions with $q=1, 5, 10$ and spacetime dimension $d=4$. The first massive mode is located on the brane in the sense that the wave function peaks on the brane only and decays when it goes far from the brane. The larger the brane tension is, the well the localization is.}
\label{firstwave4d}
\end{figure}

\begin{figure}[t]
\centering
\includegraphics[width=10cm]{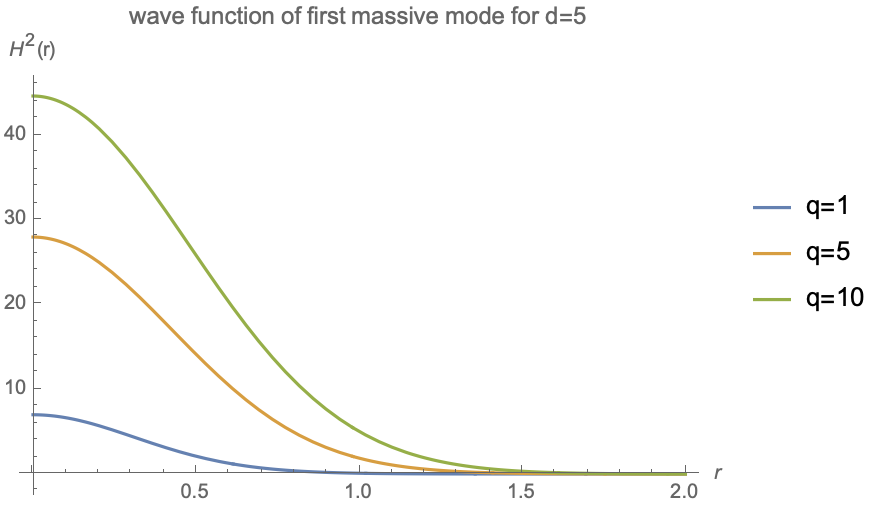}
\caption{Wave function $H^2(r)$ of the first massive gravitational mode for tensions with $q=1, 5, 10$ and spacetime dimension $d=5$. The first massive mode is located on the brane in the sense that the wave function peaks on the brane only and decays when it goes far from the brane. The larger the brane tension is, the well the localization is.}
\label{firstwave5d}
\end{figure}

\begin{figure}[t]
\centering
\includegraphics[width=7.2cm]{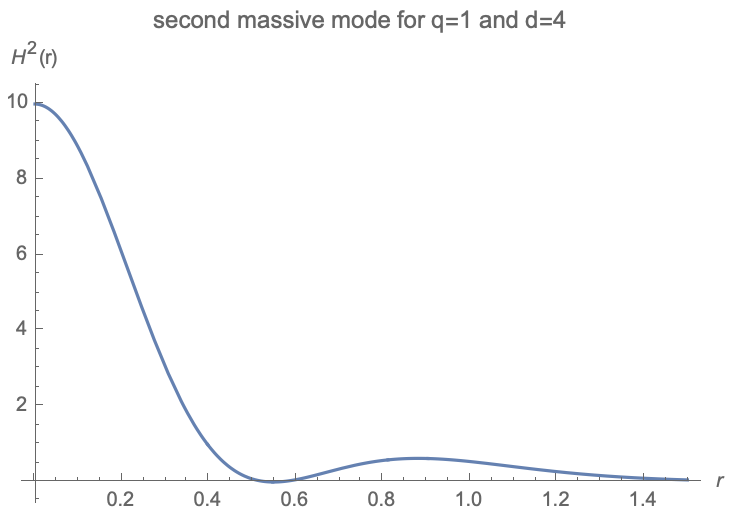}\includegraphics[width=9cm]{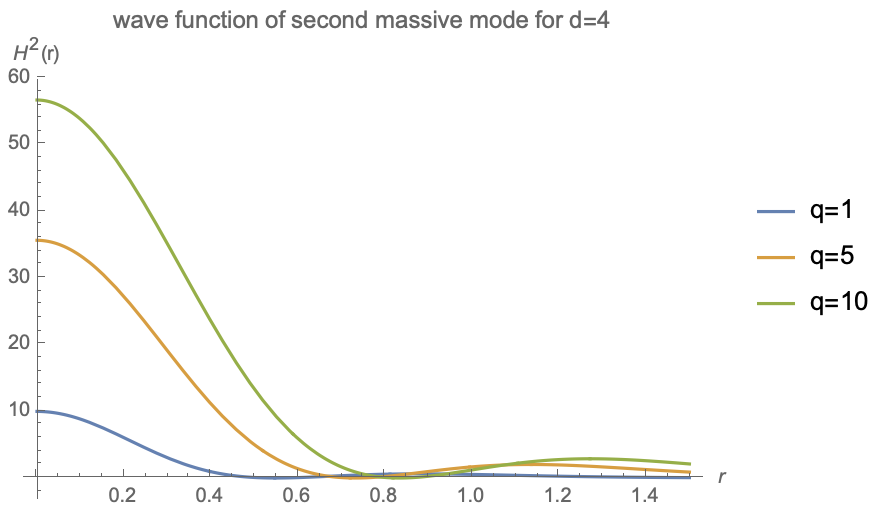}
\caption{Wave function $H^2(r)$ of the second massive gravitational mode for tensions with $q=1, 5, 10$ and spacetime dimension $d=4$.  The left figure is for $q=1$ and the right figure is for $q=1,5,10$. The wave function of second massive modes oscillate and do not decay monotonically when it goes far from the brane. Thus it is not well-located on the brane.}
\label{secondwave4d}
\end{figure}

\begin{figure}[t]
\centering
\includegraphics[width=7.2cm]{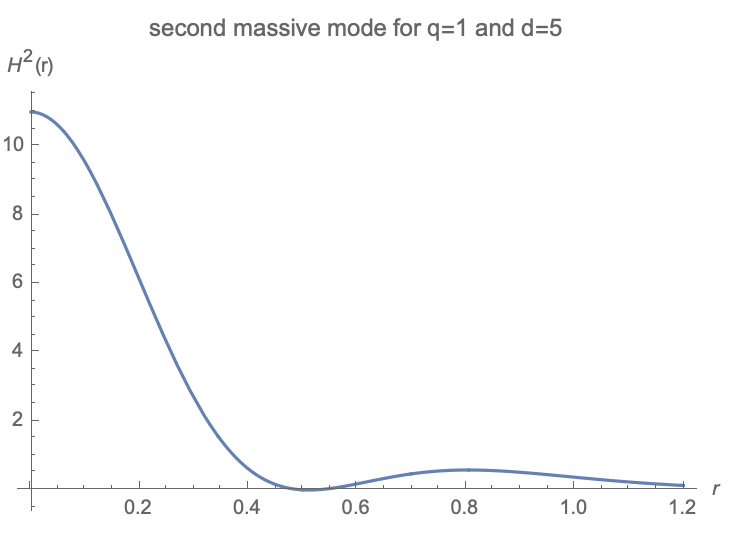}\includegraphics[width=9cm]{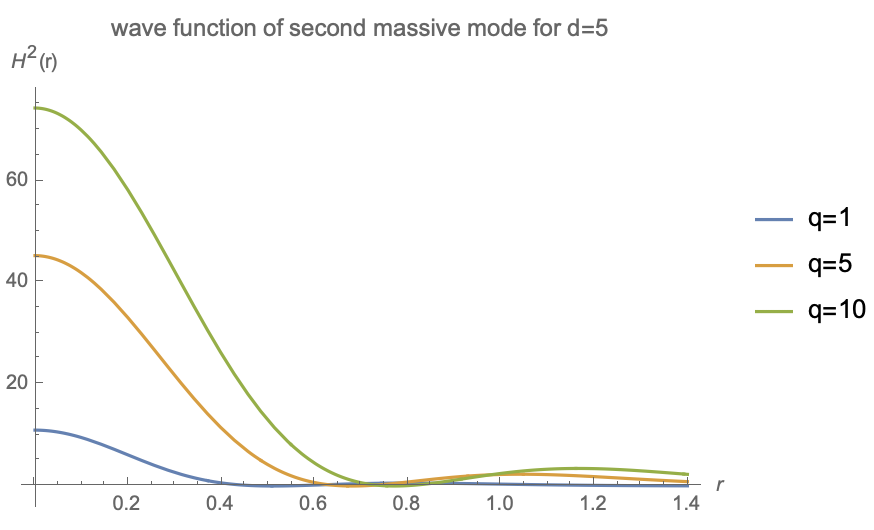}
\caption{Wave function $H^2(r)$ of the second massive gravitational mode for tensions with $q=1, 5, 10$ and spacetime dimension $d=5$.  The left figure is for $q=1$ and the right figure is for $q=1,5,10$. The wave function of second massive modes oscillate and do not decay monotonically when it goes far from the brane. Thus it is not well-located on the brane.}
\label{secondwave5d}
\end{figure}

From Fig. \ref{firstwave4d} and Fig. \ref{firstwave5d}, we see that the first massive gravitational mode is located on the brane in the sense that the wave function peaks on the brane only and decays when it goes far from the brane. The larger the brane tension is, the better the localization is. That is reasonable since heavier branes have stronger gravitational force on the fluctuation modes on the brane. On the other hand, the other massive modes oscillate and do not decay monotonically when it goes far from the brane. See Fig. \ref{secondwave4d} and Fig. \ref{secondwave5d} for the wave function of the second massive modes. Thus only the first massive gravitational mode is well-located on the brane. With a located graviton on the brane, one can study the black hole evolution and island on the codim-2 branes.

\subsubsection{Effective potential}

Let us go on to study the effective potential energy for gravitational fluctuations. We find that the effective potential takes the minimum value on the location of the brane. As a result, the gravitational fluctuations tend to be located on the brane, which is consistent with the discussions of wave functions in the above subsection. However, we do not find delta-function potential in the codim-2 brane, which is different from the ``volcanic potential" of the codim-1 brane.

Redefining 
\begin{eqnarray}\label{3.2.2:fgBA}
g(r)=\exp(2A(r)),\  f(r)=\exp(2B(r)),
\end{eqnarray}
the EOM of $H(r)$ (\ref{EOMmassiveH}) becomes
\begin{eqnarray}\label{3.2.2:BAEOMH}
H''(r)+H'(r) \left((d-1) A'(r)+B'(r)\right)+m^2 e^{-2 A(r)} H(r)=0.
\end{eqnarray}
Following \cite{Li:2020mgc}, we define the wave function
\begin{eqnarray}\label{3.2.2:wavefunction}
\Psi(r)= \exp{\Big( \frac{d-1}{2} A(r)\Big)} H(r), 
\end{eqnarray}
and rewrite EOM (\ref{3.2.2:BAEOMH}) into a Schrodinger-like equation
\begin{eqnarray}\label{3.2.2:SchrodingerEOM}
\Big(-\widetilde{\Box} +V \Big) \Psi= m^2 \Psi,\ \ V= \exp{\Big(\frac{1-d}{2} A(r)\Big)} \widetilde{\Box}  \exp{\Big( \frac{d-1}{2} A(r)\Big)},
\end{eqnarray}
where $\widetilde{\Box}$ is the Laplacian operator with respect to the metric  $\widetilde{g}_c=e^{-2A} \text{diag}(1, e^{2B})$, which is conformally equivalent to the conical metric $g_c=\text{diag}(1, e^{2B})$.   

\begin{figure}[t]
\centering
\includegraphics[width=8.5cm]{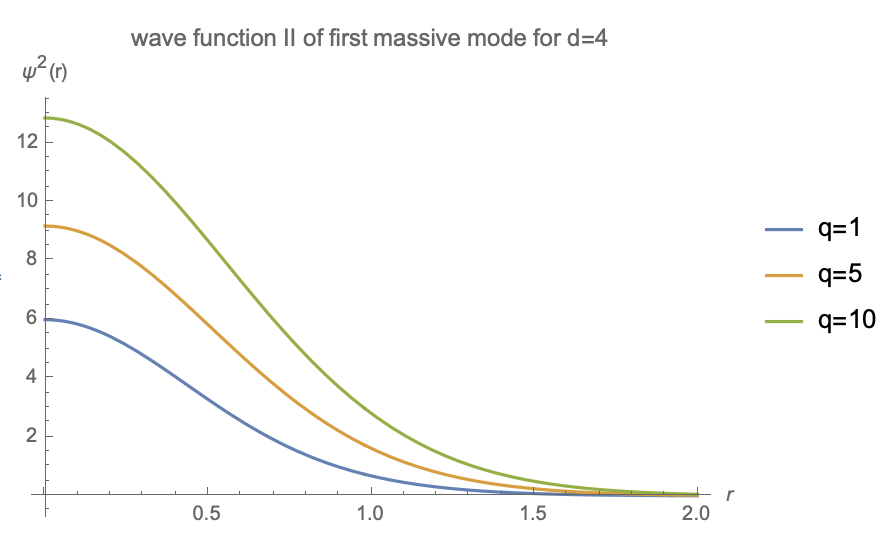}\includegraphics[width=8.5cm]{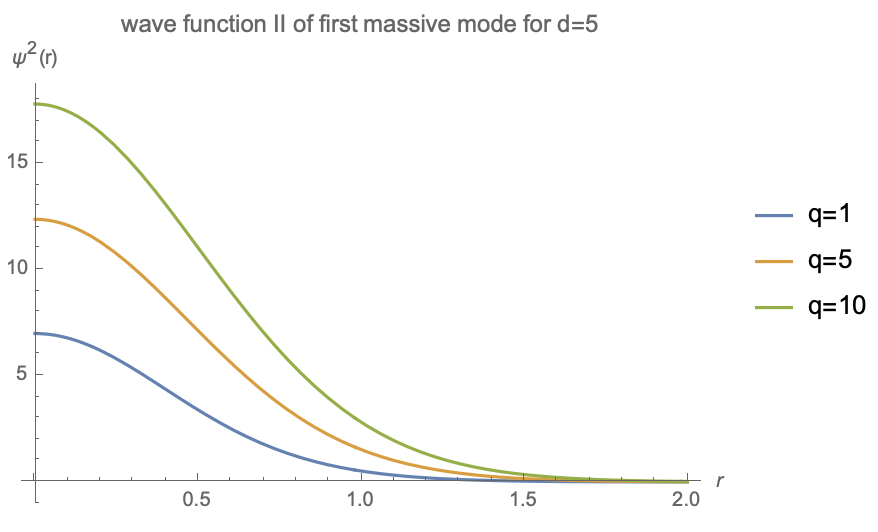}
\caption{Wave function $\Psi^2(r)$ of the first massive gravitational mode for tensions with $q=1, 5, 10$ and spacetime dimension $d=4$ (left) and $d=5$ (right). The first massive mode is located on the brane in the sense that the wave function peaks on the brane only and decays when it goes far from the brane. The larger the brane tension is, the well the localization is.}
\label{firstwaveII4d5d}
\end{figure}

Note that the shape of functions $\Psi(r)$ and $H(r)$ are very similar, thus we can take either $\Psi(r)$ and $H(r)$ as the ``wave function" \cite{Li:2020mgc}.  See Fig. \ref{firstwaveII4d5d} for wave functions $\Psi^2(r)$, whose shapes are similar to those of $H^2(r)$. 

Now let us discuss the effective potential $V$, which is given by (\ref{3.2.2:SchrodingerEOM})
\begin{eqnarray}\label{3.2.2:V}
V&=&\frac{1}{4} (d-1) e^{2 A(r)} \left(2 A''(r)+2 A'(r) B'(r)+(d-1) A'(r)^2\right)\nonumber\\
&=&\frac{(d-1) \left((d-5) f(r) g'(r)^2+2 g(r) \left(f'(r) g'(r)+2 f(r) g''(r)\right)\right)}{16 f(r) g(r)}.
\end{eqnarray}
Note that for the conical metric $g_c=\text{diag}(1, e^{2B})=\text{diag}(1, f(r) )$ or the (\ref{perturbationmetric}) with (\ref{3.2.2:fgBA}), it is $f''(r)$ instead of $g''(r)$ that contributes to the delta-function potential \cite{Bostock:2003cv}
\begin{eqnarray}\label{3.2.2:deltafunction}
\frac{d^2}{dr^2} \sqrt{f(r)}=-(1-\frac{1}{q}) \delta(r).
\end{eqnarray}
Since the effective potential includes no $f''(r)$, it includes no the delta-function potential, which is quite different from the case of codim-1 brane.  Note also that, for codim-1 brane in AdS/BCFT,  $g''(r)$ indeed yields a delta-function potential proportional to $\sinh(\rho) \delta(r+\rho)$, where EOW brane is located at $r=-\rho$ with $-\rho\le r\le \infty$. While for codim-2 brane in AdS/dCFT, we have $0\le r \le \infty$ ($\rho=0$ effectively). As a result, the potential delta function $ \lim_{\rho\to 0}\sinh(\rho) \delta(r+\rho)=0$ disappears. 

\begin{figure}[t]
\centering
\includegraphics[width=8cm]{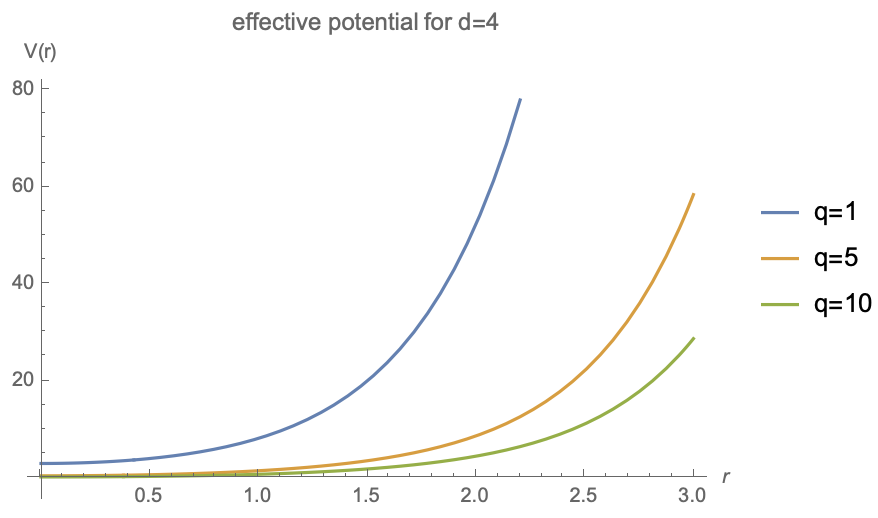}\includegraphics[width=8cm]{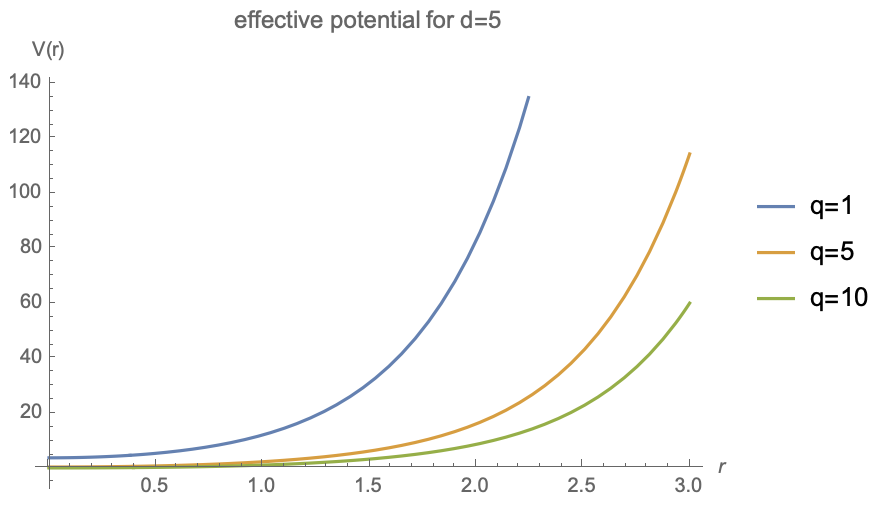}
\caption{The the effective potential for $d=4$ (left) and $d=5$ (right). The effective potential increases with $r$ and takes the minimal value on the brane $r=0$. Besides, the larger the tension 
($q$) is, the smaller the effective potential is. This implies that the gravity tends to be located on the brane. And the larger the tension is, the well the localization is.  }
\label{potential4d5d}
\end{figure}

From (\ref{f4d},\ref{g4d},\ref{3.2.2:fgBA},\ref{3.2.2:V}), we get the effective potential for $d=4$
\begin{eqnarray}\label{3.2.2:4dV}
V=\frac{3 \left(1-2 \bar{r}_h^2\right) \left((5 \cosh (4 r)+11) \bar{r}_h^2+2 \sinh ^2(r) (3-5 \cosh (2 r))\right)}{16 \left(\sinh ^2(r)-\cosh (2 r) \bar{r}_h^2\right)}.
\end{eqnarray}
As for the case of $d>4$,  we have to do numerical calculations. See Fig. \ref{potential4d5d} for examples. It is shown that the effective potential increases with $r$ and takes the minimal value on the brane $r=0$. Besides, the larger the tension 
($q$) is, the smaller the effective potential is. This implies that the gravity tends to be located on the brane. And the larger the tension is, the better the localization is.

To summarize, by studying the wave function and the effective potential, we find that the first massive gravitational mode is well located on the codim-2 brane. Finally, we want to mention that the massive gravitational modes are normalizable (\ref{3.1:normalizedH}), which also supports the localization of gravitations on the brane.

\section{Page curve on codim-2 brane in $\text{AdS}_4/\text{dCFT}_3$}

In this section, we study the island and Page curve on codim-2 branes in AdS/dCFT. To warm up, we first study the case of $\text{AdS}_4/\text{dCFT}_3$.  Compared with $\text{AdS}/\text{dCFT}$ in higher dimensions, more analytical results can be obtained in this toy model. As expected, the Page curve of eternal black holes can be recovered due to the island on the codim-2 brane. However, the extremal surface passing through the horizon cannot be defined after some finite time, which is quite different from the case of codim-1 brane in AdS/BCFT. Fortunately, this unusual situation happens only after Page time. 
As a result, it does not affect the Page curve.  For simplicity, we focus on the tensionless brane below and comment on the tensive case at the end of this section. 

We take the following ansatz of the bulk metric
\begin{eqnarray}\label{sect4:metrictoymodel}
ds^2=dr^2+\sinh^2(r) d\theta^2+ \cosh^2(r) \frac{\frac{dz^2}{1-z^2}-(1-z^2) dt^2}{z^2},
\end{eqnarray}
where the codim-2 brane is located at $r=0$ and the bath is at the AdS boundary $r\to \infty$.  There is a horizon located at $z=1$ on both the codim-2 brane and the AdS boundary. Note that the codim-2 brane in $\text{AdS}_4/\text{dCFT}_3$ is two dimensional. Similar to the case of codim-1 brane in  $\text{AdS}_3/\text{BCFT}_2$, it is expected that the effective theory on the codim-2 brane in $\text{AdS}_4/\text{dCFT}_3$ is JT gravity. How to derive JT gravity on the codim-2 brane is beyond the purpose of this paper, and we leave this interesting problem to future works. 

Following \cite{Almheiri:2019yqk,Almheiri:2019psy}, we focus on the eternal two-sided black hole, which is dual to the thermofield double state of CFTs \cite{Eternal black hole M}
\begin{eqnarray}\label{TFD}
| \text{TFD}\rangle=Z^{-1/2} \sum_{\alpha} e^{-E_{\a}/(2T)}e^{-iE_{\a}(t_L+t_R)} | E_{\a}\rangle_L  | E_{\a}\rangle_R,
\end{eqnarray}
where $L$ and $R$ label the states (times) associated with the left and right boundaries.  Couple the eternal two-side black hole to bathes on each 
side. The system evolutes if we move time $t_L$ and $t_R$ forward on both sides.  Without the island, the entanglement entropy will increase with time and exceed the double black hole entropy in the late times. However, the fine-grained entanglement entropy cannot be larger than the coarse-grained black hole entropy \cite{Almheiri:2020cfm}. This leads to the information paradox for eternal black holes.  Thanks to the island outside the horizon, the entanglement entropy becomes a constant at Page time, which is smaller than the double black hole entropy. In this way, the information paradox of eternal black holes can be resolved. 

\begin{figure}[!ht]
  \centering
   \begin{tikzpicture} [scale=5]
\draw[dashed,ultra thick,black] ({1},{0}) arc (0:90:{1});
\draw[ultra thick,lime!80!black] (0.8,0)  to [out=90,in=285] (0.7,{sqrt(1-0.7*0.7)});
\draw[-,ultra thick,,cyan!60!black] ({0.8},{0}) arc (0:90:{0.8});
\draw[-,ultra thick,yellow!90!black] (0,0) -- (0.8,0) ;
\draw[-,ultra thick,purple!70!black] (0.8,0) -- (1,0) ;
\draw[-,ultra thick,orange!95!black] (0,0) -- (0,0.8) ;
\draw[-,ultra thick,magenta!80!white] (0,0.8) -- (0,1) ;
\fill [black] (0,0) circle [radius=0.015];
\draw[-stealth] (-0.15,0.4) -- (-0.15,0.6) ;
\node[left] at  (-0.15,0.5)  {$z$};
\draw[-stealth] ({1.15*cos(50)},{1.15*sin(50)}) arc (50:40:{1.15});
\node[above,right] at  ({0.80},{0.85})  {$r$};
\node[right] at  ({1},{0})  {$r\to \infty$};
\node[above] at  ({0},{1})  {$r=0$};
\node[below] at  (0,0)  {Defect};
\node[below] at  (0.4,0)  {$\bar{R}$};
\node[below] at  (0.9,0)  {${R}$};
\node[left] at  (0,0.4)  {$\bar{I}$};
\node[left] at  (0,0.9)  {${I}$};
\draw [decorate,decoration={text along path,text align=center,text={|\fontsize{5}{5}\selectfont| Horizon}}] ({0},{1.02}) arc (90:20:{1.02});
\draw [decorate,decoration={text along path,text align=center,text={|\fontsize{5}{5}\selectfont| Island Surface}}] ({0},{0.82}) arc (90:20:{0.82});
\draw [decorate,decoration={text along path,text align=center,text={|\fontsize{5}{5}\selectfont| No-island Surface}}]  (0.72,{sqrt(1-0.72*0.72)}) to [in=90,out=285] (0.815,0.1) ;
\node[above,rotate=270] at  (0,0.9)  {{\fontsize{5}{5}\selectfont Island}};
\node[above,rotate=270] at  (0,0.4)  {{\fontsize{5}{5}\selectfont Island Complement}};
\node[above] at  (0.4,0)  {{\fontsize{5}{5}\selectfont Radiation Complement}};
\node[above] at  (0.9,0.05)  {{\fontsize{4}{4}\selectfont Radiation}};
\node[above] at  (0.9,0)  {{\fontsize{4}{4}\selectfont Region}};
\end{tikzpicture}
   \begin{tikzpicture} [scale=5]
 \draw[dashed,ultra thick,black] ({1},{0}) arc (0:90:{1});
\draw[ultra thick,lime!80!black] (0.8,0)  to [out=90,in=280] (0.75,{sqrt(1.1-0.75*0.75)});
\draw[-,ultra thick,,cyan!60!black] ({0.8},{0}) arc (0:90:{0.8});
\draw[-,ultra thick,yellow!90!black] (0,0) -- (0.8,0) ;
\draw[-,ultra thick,purple!70!black] (0.8,0) -- (1,0) ;
\draw[-,ultra thick,orange!95!black] (0,0) -- (0,0.8) ;
\draw[-,ultra thick,magenta!80!white] (0,0.8) -- (0,1) ;
\fill [black] (0,0) circle [radius=0.015];
\draw[-stealth] (-0.15,0.4) -- (-0.15,0.6) ;
\node[left] at  (-0.15,0.5)  {$z$};
\draw[-stealth] ({1.15*cos(50)},{1.15*sin(50)}) arc (50:40:{1.15});
\node[above,right] at  ({0.80},{0.85})  {$r$};
\node[right] at  ({1},{0})  {$r\to \infty$};
\node[above] at  ({0},{1})  {$r=0$};
\node[below] at  (0,0)  {Defect};
\node[below] at  (0.4,0)  {$\bar{R}$};
\node[below] at  (0.9,0)  {${R}$};
\node[left] at  (0,0.4)  {$\bar{I}$};
\node[left] at  (0,0.9)  {${I}$};
\draw [decorate,decoration={text along path,text align=center,text={|\fontsize{5}{5}\selectfont| Horizon}}] ({0},{1.02}) arc (90:20:{1.02});
\draw [decorate,decoration={text along path,text align=center,text={|\fontsize{5}{5}\selectfont| Island Surface}}] ({0},{0.82}) arc (90:20:{0.82});
\draw [decorate,decoration={text along path,text align=center,text={|\fontsize{5}{5}\selectfont| No-island Surface}}]  (0.76,{sqrt(1.05-0.75*0.75)}) to [in=90,out=285] (0.82,0) ;
\node[above,rotate=270] at  (0,0.9)  {{\fontsize{5}{5}\selectfont Island}};
\node[above,rotate=270] at  (0,0.4)  {{\fontsize{5}{5}\selectfont Island Complement}};
\node[above] at  (0.4,0)  {{\fontsize{5}{5}\selectfont Radiation Complement}};
\node[above] at  (0.9,0.05)  {{\fontsize{4}{4}\selectfont Radiation}};
\node[above] at  (0.9,0)  {{\fontsize{4}{4}\selectfont Region}};
\end{tikzpicture}
\caption{The geometry of AdS/dCFT with a constant $\theta$ and its physical interpretation in the black hole information paradox. The left figure is for $t=0$ and the right figure is for $t>0$. The black hole lives on codim-2 brane $E$ at $r=0$, and CFT bath lives on the AdS boundary $M$ at $r\to \infty$. The island, island complement, radiation complement, and radiation are denoted by the red, orange, yellow, and purple segments, respectively. The blue line denotes the extremal surface in the island phase, and the green line is the extremal surface in the no-island phase. The extremal surface in the no-island phase (green line) ends on the horizon (dotted line) at the beginning $t=0$ and passes through the horizon at $t>0$. Holographic entanglement entropy is dominated by the green extremal surface (no-island phase) at early times and the blue extremal surface (island phase) at late times.}
  \label{fig:doublyholographicsetup}
\end{figure}
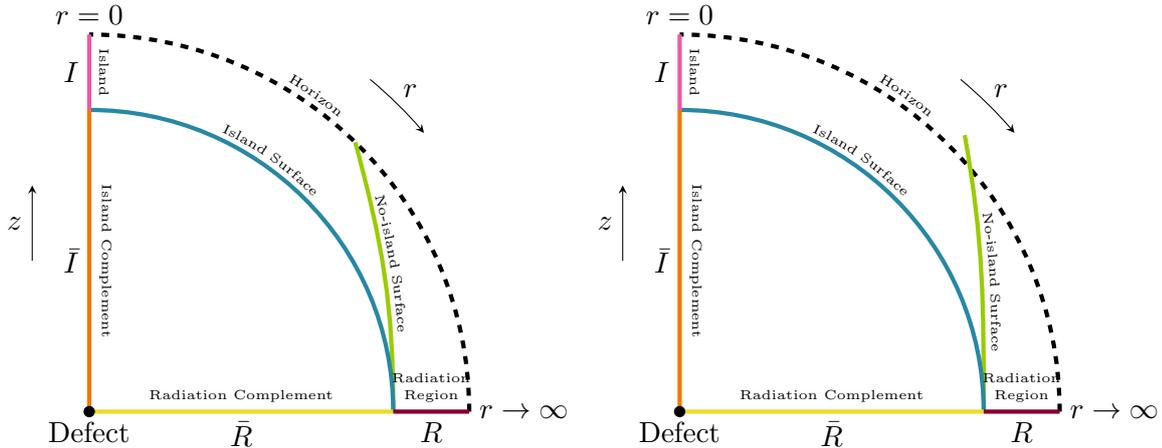

Following the convention of \cite{Geng:2021mic},  let us illustrate the geometry of AdS/dCFT and its physical interpretation in the black hole information paradox in Fig.\ref{fig:doublyholographicsetup}.  Recall that the geometry (\ref{BHmetricr}) of AdS/dCFT is axisymmetric, i.e., $\theta \simeq \theta + 2\pi q$. Without loss of generality, we focus on the constant $\theta$,  then the geometry reduces to that of black string in AdS/BCFT \cite{Geng:2021mic}. For simplicity, we show only one side of the black holes in the figure. One can double the figure for the two-side black holes as in \cite{Geng:2021mic}.  As shown in Fig.\ref{fig:doublyholographicsetup}, the black hole lives on codim-2 brane $E$ at $r=0$, and CFT bath lives on the AdS boundary $M$ at $r\to \infty$. The island, island complement, radiation complement, and radiation regions are denoted by the red, orange, yellow, and purple segments, respectively. The blue line denotes the extremal surface in the island phase, and the green line is the extremal surface in the no-island phase. The extremal surface in the no-island phase (green line) ends on the horizon (dotted line) at the beginning $t=0$ and passes through the horizon at $t>0$. Holographic entanglement entropy is dominated by the green extremal surface (no-island phase) at early times and the blue extremal surface (island phase) at late times.

It should be mentioned that the Ryu-Takayanagi formula \cite{Ryu:2006bv} still applies to AdS/dCFT. Using the approach of Casini, Huerta, and Myers \cite{Casini:2011kv}, \cite{Jensen:2013lxa} proves the Ryu-Takayanagi formula for the spherical entangling surface centered on the codim-n defect. See also \cite{Kobayashi:2018lil}. This is precisely the case we focus on in this paper. As for the entangling surfaces with general shapes, one can use the method of Lewkowycz and Maldacena \cite{Lewkowycz:2013nqa} to derive the Ryu-Takayanagi formula. However, this is beyond the primary purpose of this paper, and we leave it to future works. Finally, we want to mention that, as a minimal surface, the Ryu-Takayanagi (RT) surface should be perpendicular to the codim-2 brane in the island phase.

\subsection{Island phase}

Let us first study the island phase, where the RT surface ends on the brane.  The embedding function of extremal surfaces are given by
\begin{eqnarray}\label{sect4.1:embedding function}
t=\text{constant}, \ \ z=z(r).
\end{eqnarray}
Substituting (\ref{sect4.1:embedding function}) into (\ref{sect4:metrictoymodel}), we get the area of RT surface
\begin{eqnarray}\label{sect4.1:island area function} 
A=2\pi\int^{r_{\text{UV}}}  \sinh (r) \sqrt{1+\frac{\cosh ^2(r) z'(r)^2}{z(r)^2-z(r)^4}} \ dr, 
\end{eqnarray} 
where $r_{\text{UV}}$ is a UV cut-off. 
Note that (\ref{sect4.1:island area function}) is the area of RT surface in the one-side black hole. For the two-side black hole, we double the above result. We take this notation in the following of this paper.

\begin{figure}[t]
\centering
\includegraphics[width=11cm]{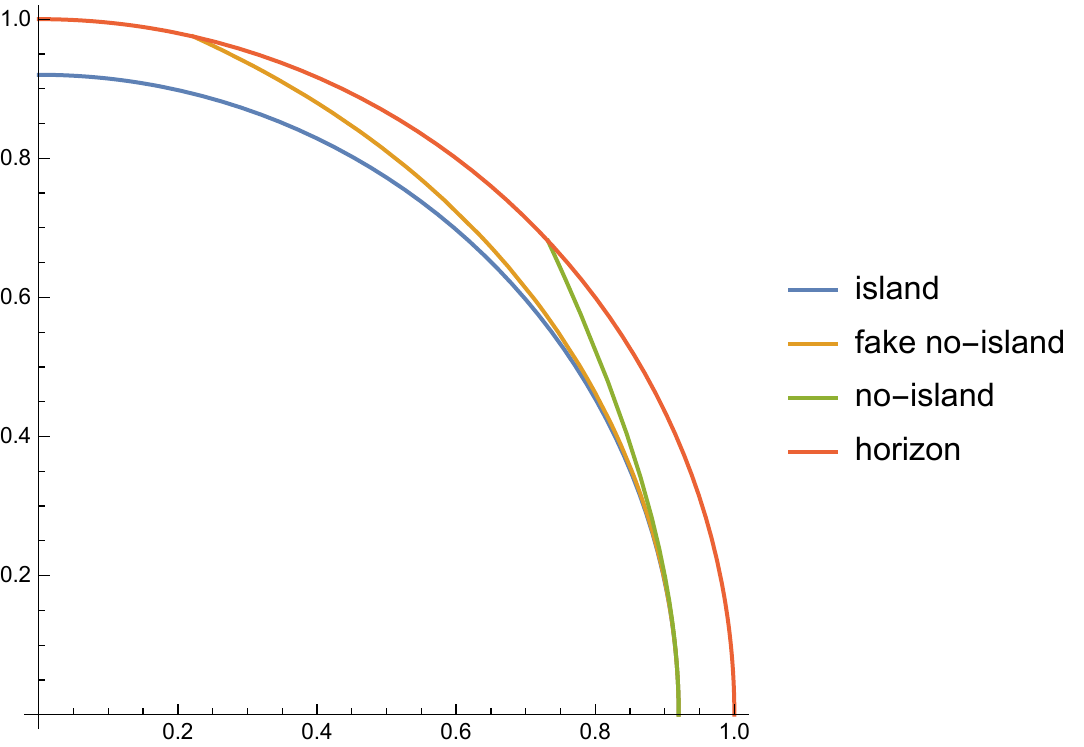}
\caption{RT surfaces in the polar coordinates $(z, \phi)$ at $t_R=t_L=0$, where $\phi=2 \tan ^{-1}\left(e^{-r}\right)$ and we have chosen 
$z_{\text{bdy}}=0.92$. The brane is at $\phi=\pi/2$, the AdS boundary is at $\phi=0$ and the horizon (red line) is at $z=1$. The RT surface in the island phase is labelled by the blue line, which is perpendicular to both the AdS boundary and the brane. The RT surface in the no-island phase is labelled by the green line, which is perpendicular to both the AdS boundary and the horizon. The orange line denotes the fake RT surface in the no-island phase, which does not contribute to the entanglement entropy since it has larger area than the green line. 
Note that we are discussing the AdS space instead of the flat space. Although it appears not to be, the green and orange lines are perpendicular to the horizon (red line). }
\label{RTsurfaces}
\end{figure}

 Taking variations of (\ref{sect4.1:island area function}), we get the equation of motion (EOM)
\begin{eqnarray}\label{sect4.1:EOMZ} 
&&z(r) z'(r) \left(\left(1-2 z(r)^2\right) z'(r)+z(r) \tanh (r) \left(\coth ^2(r)+2\right) \left(z(r)^2-1\right)\right)\nonumber\\
&&+z(r)^2 \left(z(r)^2-1\right) z''(r)- z'(r)^3 (\sinh (2 r)+\coth (r))=0.
\end{eqnarray}
Interestingly, there is an exact solution to EOM (\ref{sect4.1:EOMZ})
\begin{eqnarray}\label{sect4.1:solution1} 
z(r)=z_{\text{bdy}},
\end{eqnarray} 
where $0\le z_{\text{bdy}}\le 1$ is the value of $z$ on the AdS boundary $r\to \infty$. (\ref{sect4.1:solution1}) is the solution in the island phase, where the RT surface is perpendicular to both the AdS boundary $r\to \infty$ and the codim-2 brane $r=0$. 
See blue line of Fig.\ref{RTsurfaces}. Note that we require $0\le z_{\text{bdy}}\le 1$ so that both the island and the bath are outside the horizon $z=1$. 
For simplicity, we choose $z_{\text{bdy}}$ as a free parameter in this paper. Similar to the case of codim-1 brane, in principle, $z_{\text{bdy}}$ can be fixed by 
considering suitable DGP terms for $d>3$ (JT gravity for $d=3$) or matter fields on the codim-2 brane and then minimizing entanglement entropy. See \cite{Ling:2020laa,Chen:2020hmv} for examples. Substituting (\ref{sect4.1:solution1}) into (\ref{sect4.1:island area function}), we derive the area of RT surface in the island phase
\begin{eqnarray}\label{sect4.1:island area function1} 
A_{\text{island}}=2\pi \int_{0}^{r_{\text{UV}}} \sinh (r) dr=2\pi \big( \cosh(r_{\text{UV}})-1 \big).
\end{eqnarray} 

\begin{figure}[t]
\centering
\includegraphics[width=8cm]{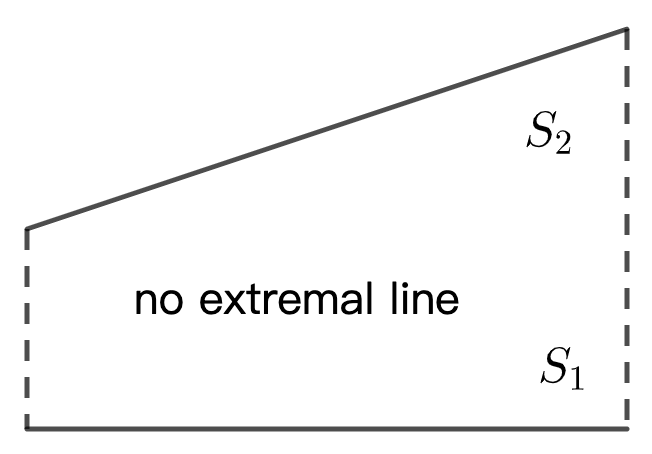}\includegraphics[width=8cm]{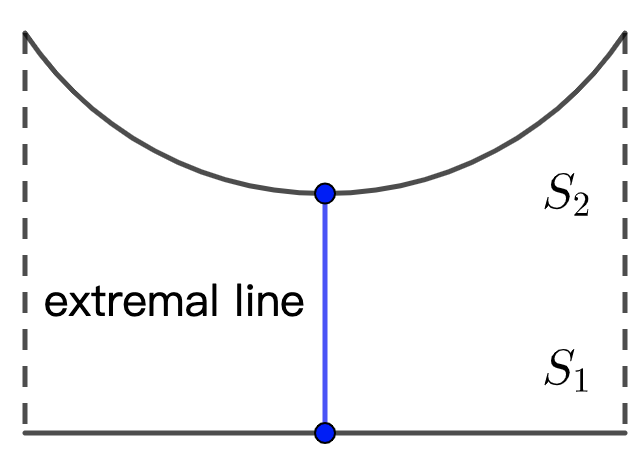}
\caption{ The extremal lines perpendicular to two boundaries $S_1$ and $S_2$ in a flat plane. (Left figure) $S_1$ and $S_2$ are two non-parallel line segments, and there is no extremal line perpendicular to both $S_1$ and $S_2$. (Right figure) $S_1$ is a line segment and $S_2$ is a circular arc. There is only one extremal line (blue line) perpendicular to both $S_1$ and $S_2$. It is shown that the vertical condition to the boundaries highly restricts the extremal lines. In particular, the allowed ending points of extremal lines on the boundary $S_1$ are limited, which is similar to the case of codim-2 branes 
in AdS. }
\label{vertical condition}
\end{figure}

In the above discussion, we focus on a particular solution to (\ref{sect4.1:EOMZ}), which is in the island phase. 
Let us go on to discuss the solution in the no-island phase at $t_R=t_L=0$. Unfortunately, there is no exact expression for this kind of solution. Thus, we have to do numerical calculations.  As a minimal surface, the RT surface in no-island phase should be perpendicular to both the AdS boundary $r\to \infty$ and the black hole horizon $z=1$ at $t_R=t_L=0$. This is a highly non-trivial boundary condition, which significantly restricts the regions of ending points of RT surfaces. Let us study some examples in a flat space to get some feelings 
on this point.  See Fig. \ref{vertical condition}, where we want to find the extremal line perpendicular to two boundaries $S_1$ and $S_2$ in a flat plane. As shown in Fig. \ref{vertical condition}, the allowed ending points of extremal lines on the boundary are limited. This is similar to the case of codim-2 branes in an AdS space, which will be discussed below. 

Solving EOM (\ref{sect4.1:EOMZ}) perturbatively around the horizon $z=1$, we get
\begin{eqnarray}\label{sect4.1:Znearhorizon} 
z(r)=1+ a_1 (r-r_0)+ a_2 (r-r_0)^2+ O(r-r_0)^3,
\end{eqnarray} 
where $r_0$ is value of $r$ on horizon and $a_i$ are given by
\begin{eqnarray}\label{sect4.1:Za1} 
&&a_1=-4 \sinh ^2\left(r_0\right) \text{csch}\left(4 r_0\right),\\
&& a_2=\frac{1}{6} \left(8 \cosh \left(2 r_0\right)-7\right) \text{sech}^2\left(2 r_0\right). \label{sect4.1:Za2} 
\end{eqnarray}
Note that we have selected the non-constant solution for $z(r)$ above. One can check that the solution (\ref{sect4.1:Znearhorizon},\ref{sect4.1:Za1},\ref{sect4.1:Za2}) obey the vertical condition that the RT surface is perpendicular to the horizon.  
\begin{figure}[t]
\centering
\includegraphics[width=10cm]{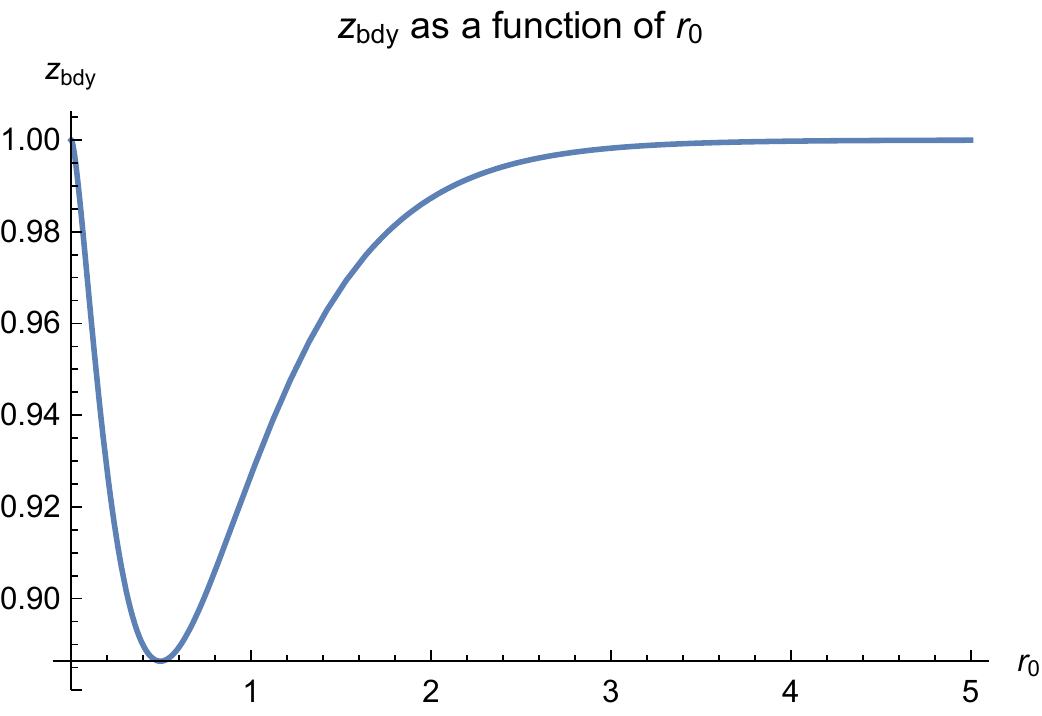}
\caption{$z_{\text{bdy}}=z(r_\text{UV})$ as a function of $r_0$, where $r_0$ is the value of $r$ on horizon. It is shown that there is a lower bound of $z_{\text{bdy}}\ge z_c \approx 0.886$. The  corresponding critical value of $r_0$ is $r_c\approx 0.496$. Besides, 
one $z_{\text{bdy}}$ corresponds to two $r_0$ for $z_{\text{bdy}}\ge z_c$. It turns out that the extremal surface with larger $r_0$ has smaller area. }
\label{lower bound of zbdy}
\end{figure}

For any given $0\le r_0 \le r_{\text{UV}}$, we can numerically solve EOM (\ref{sect4.1:EOMZ}) together with BC (\ref{sect4.1:Znearhorizon}) and then obtain $z_{\text{bdy}}=z(r_{\text{UV}})$ on the AdS boundary. It turns out that there is a lower bound of $z_{\text{bdy}}$,
which agrees with the above discussions that the vertical BC highly restricts the locations of ending points of extremal surfaces on the boundary.  See Fig. \ref{lower bound of zbdy} for more details, where the minimal value of $z_{\text{bdy}}$ is approximately by $0.886$
\begin{eqnarray}\label{sect4.1:bound of zbdy} 
z_{\text{bdy}}\ge z_c\approx 0.886,
\end{eqnarray}
at $r_0\approx 0.496$. Besides, one $z_{\text{bdy}}$ corresponds to two $r_0$ for $z_{\text{bdy}}> z_c$ . It turns out that the extremal surface with larger $r_0$ has smaller area.  Take 
$z_{\text{bdy}}=0.92$ with $r_0\approx 0.226$ and $r_0\approx 0.933$ as an example, we have 
\begin{eqnarray}\label{sect4.1:twoarea} 
A_{\text{no-island}}-A_{\text{island}}
&\approx&\begin{cases}
0.214 ,
 \ \ \ \text{for} \ r_0\approx 0.226,\\
-0.223,
\ \text{for} \ r_0\approx 0.933,
\end{cases}
\end{eqnarray}
where $A_{\text{no-island}}$ is the area of RT surface at the beginning $t_R=t_L=0$ in the no-island phase
\begin{eqnarray} \label{sect4.1:noisalnd} 
A_{\text{no-island}}=2\pi\int_{r_0}^{r_{\text{UV}}} \sinh (r) \sqrt{1+\frac{\cosh ^2(r) z'(r)^2}{z(r)^2-z(r)^4}} dr.
\end{eqnarray}
Since the holographic entanglement entropy is related to the extremal surface with smaller area, we always select the one with larger $r_0$ in the followings of this paper.  The extremal surfaces associated to $r_0\approx 0.226$ and $r_0\approx 0.933$ are labelled by the orange line and the 
green line of Fig.\ref{RTsurfaces}, respectively. It should be mentioned that the existence of a lower bound (\ref{sect4.1:bound of zbdy}) of $z_{\text{bdy}}$ is a new feature for codim-2 branes, and there is no such lower bound for codim-1 branes.

To have the no-island phase, we need $A_{\text{non-island}}< A_{\text{island}}$ at the beginning $t_R=t_L=0$. This leads to another lower bound of $z_{\text{bdy}}$,
\begin{eqnarray}\label{sect4.1:bound of zbdy1} 
z_{\text{bdy}}> \bar{z}_c\approx 0.911,
\end{eqnarray}
around $r_0\approx 0.847$.  To summarize, we take the following parameter space in this section 
\begin{eqnarray}\label{sect4.1:bound of zbdy2} 
0.911 < z_{\text{bdy}} < 1,\ \ 0.847 < r_0 < r_\text{UV}. 
\end{eqnarray}

\subsection{No-island phase}

Let us go on to study the RT surface passing through the horizon in general times 
$(t_R=t_L\ge 0)$, 
which is the so-called no-island phase. The area of this kind of RT surface increases with time due to the dynamic nature of spacetime inside the horizon. Unlike the case of codim-1 brane, the RT surface passing through the horizon cannot be defined after some finite time in the case of codim-2 branes. Fortunately, this unusual situation 
does not affect the Page curve, since it happens after Page time. 

To study the RT surface passing through the horizon, it is convenient to use the infalling Eddington-Finkelstein coordinates 
\begin{eqnarray} \label{sect4.2:coordinatev}
v=t-\int_0^z \frac{1}{1-z^2} dz=t-\frac{1}{2}\log\left|\frac{1+z}{1-z} \right|. 
\end{eqnarray}
Then the metric (\ref{sect4:metrictoymodel}) becomes
\begin{eqnarray} \label{sect4.2:metricv}
ds^2 = dr^2+\sinh^2(r) d\theta^2+ \cosh^2(r)\frac{-(1-z^2)dv^2 -2dvdz}{ z^2}.
\end{eqnarray}
We take  following ansatz of the embedding function of extremal surfaces 
\begin{eqnarray}\label{sect4.2:embedding function}
 r=r(z),\ \ v=v(z).
\end{eqnarray}
Then the area of RT surface becomes
\begin{eqnarray}\label{sect4.2:noisland area function} 
A=2\pi\int^{z_{\text{max}}}_{z_{\text{bdy}}} \sinh (r(z)) \sqrt{r'(z)^2+\frac{\cosh ^2(r(z)) v'(z) \left(\left(z^2-1\right) v'(z)-2\right)}{z^2}} \ dz,
\end{eqnarray} 
where $z_{\text{max}}\ge 1$ is the turning point 
of the two-side black hole, which obeys the condition \cite{Carmi:2017jqz}
\begin{eqnarray}\label{sect4.2:zmaxBC} 
v(z_{\text{max}})+\frac{1}{2}\log\frac{z_{\text{max}}+1}{z_{\text{max}}-1}=t(z_{\text{max}})=0,\ \ v'(z_{\text{max}})=t'(z_{\text{max}})=-\infty.
\end{eqnarray} 
Taking variations of (\ref{sect4.2:noisland area function}), we get two independent equations
\begin{eqnarray}\label{sect4.2:EOMR} 
r''(z)&=&\frac{r'(z) \left(\left(z^2-1\right) v'(z)^2-2 v'(z)-2\right)}{z}\nonumber\\
&+&\frac{\cosh (2 r(z)) \coth (r(z)) v'(z) \left(\left(z^2-1\right) v'(z)-2\right)}{z^2}\nonumber\\
&+&\frac{1}{2} r'(z)^2 (3 \cosh (2 r(z))-1) \text{csch}(r(z)) \text{sech}(r(z)),
\end{eqnarray} 
and 
\begin{eqnarray}\label{sect4.2:EOMV} 
v''(z)= \frac{v'(z) \left(\left(z^2-1\right) v'(z)^2-3 v'(z)-2\right)}{z}.
\end{eqnarray} 
Interestingly, $v(z)$ decouples with $r(z)$ in (\ref{sect4.2:EOMV}). 
Solving (\ref{sect4.2:EOMV}) with BC (\ref{sect4.2:zmaxBC}), we obtain
\begin{eqnarray}\label{sect4.2:V} 
v(z)=\frac{1}{2} \log \left|\frac{\sqrt{\frac{z_{\max }^2-z^2}{z_{\max }^2-1}}+1}{\sqrt{\frac{z_{\max }^2-z^2}{z_{\max }^2-1}}-1}\right|-\frac{1}{2} \log \left|\frac{1+z}{1-z}\right|,
\end{eqnarray} 
which is smooth on the horizon $z=1$.  From (\ref{sect4.2:coordinatev}) and (\ref{sect4.2:V}), we get
\begin{eqnarray}\label{sect4.2:T} 
t(z)=\frac{1}{2} \log \left|\frac{\sqrt{\frac{z_{\max }^2-z^2}{z_{\max }^2-1}}+1}{\sqrt{\frac{z_{\max }^2-z^2}{z_{\max }^2-1}}-1}\right|.\end{eqnarray} 
The time on the AdS boundary is defined by
\begin{eqnarray}\label{sect4.2:tRtL} 
t_R=t_L=t(z_{\text{bdy}})=\frac{1}{2} \log \left(\frac{\sqrt{\frac{z_{\max }^2-z_{\text{bdy}}^2}{z_{\max }^2-1}}+1}{\sqrt{\frac{z_{\max }^2-z_{\text{bdy}}^2}{z_{\max }^2-1}}-1}\right),
\end{eqnarray} 
which yields 
\begin{eqnarray}\label{sect4.2:zmaxtR} 
z_{\text{max}}=\sqrt{\cosh ^2\left(t_R\right)-z_{\text{bdy}}^2 \sinh ^2\left(t_R\right)}.
\end{eqnarray} 
Note that we have $z_{\text{max}}=1$ at the beginning $t_R=t_L=0$. This case has been discussed at the end of sect.4.1.

Substituting (\ref{sect4.2:V}) into (\ref{sect4.2:noisland area function},\ref{sect4.2:EOMR}) and transforming $r(z)$ into $z(r)$, we derive the area functional 
\begin{eqnarray}\label{sect4.2: area function zr} 
A_{\text{no-island}}=2\pi \int_{r_0}^{r_{\text{UV}}} \sinh (r) \sqrt{1+\frac{z_{\max }^2 \cosh ^2(r) z'(r)^2}{z(r)^2 \left(z_{\max }^2-z(r)^2\right)}} \ dr,
\end{eqnarray}
and EOM of $z(r)$
\begin{eqnarray}\label{sect4.2: EOMzr} 
&&z(r) z'(r) \left(\left(z_{\max }^2-2 z(r)^2\right) z'(r)+z(r) \tanh (r) \left(\coth ^2(r)+2\right) \left(z(r)^2-z_{\max }^2\right)\right)\nonumber\\
&&+z(r)^2 \left(z(r)^2-z_{\max }^2\right) z''(r)-z_{\max }^2 z'(r)^3 (\sinh (2 r)+\coth (r))=0,
\end{eqnarray}
which agrees with results of no-island phase (\ref{sect4.1:island area function}, \ref{sect4.1:EOMZ}) at the beginning $t_R=t_L=0$, or equivalently, $z_{\text{max}}=1$. This can be regarded as a check of our calculations.  Note that $r_0$ of (\ref{sect4.2: area function zr}) is the value of $r$ at the turning point $z=z_{\text{max}}$.  One can check that (\ref{sect4.2: EOMzr}) can also be derived by taking variations of the area functional (\ref{sect4.2: area function zr}).  This is also a double check of our results. 

\begin{figure}[t]
\centering
\includegraphics[width=15cm]{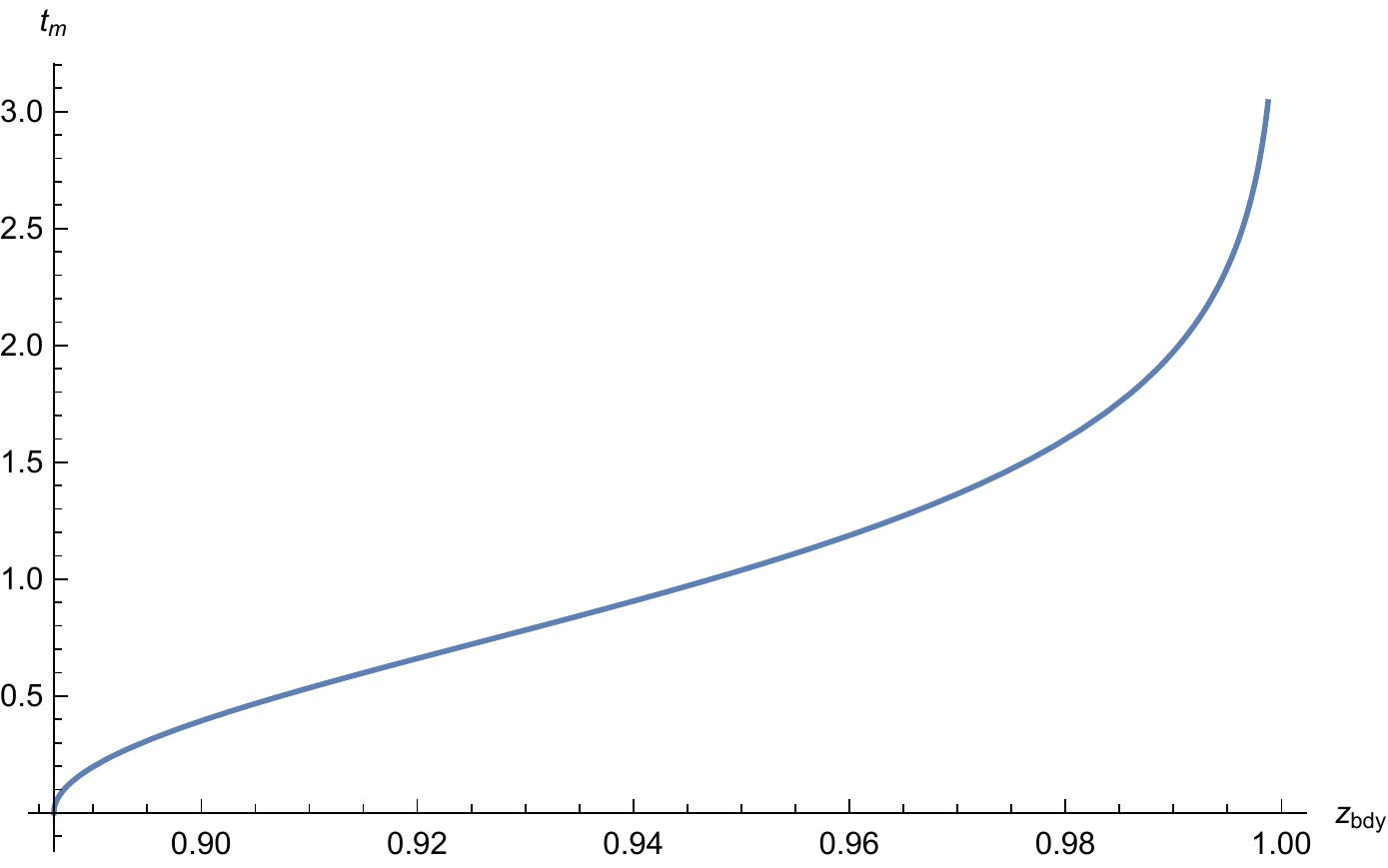}
\caption{The maximum time $t_m$ in the no-island phase increases with $z_{\text{bdy}}$.  There is an upper bound of $t_m$ for any given $z_{\text{bdy}} $ obeying $z_c\approx 0.886\le z_{\text{bdy}} < 1$. Only in the case $z_{\text{bdy}}=1$, $t_m$ can be infinity. }
\label{tmzbdy}
\end{figure}

Now the complicated time-dependent problem (\ref{sect4.2:noisland area function}) has been transformed into a simple time-independent problem (\ref{sect4.2: area function zr}), which has already been solved in sect.4.1.  To see this clearly, we make a further transformation
\begin{eqnarray}\label{sect4.2: z to Z} 
z(r)= z_{\text{max}} Z(r).
\end{eqnarray}
Then the area functional (\ref{sect4.2: area function zr}) become 
\begin{eqnarray}\label{sect4.2: area function Zr} 
A_{\text{no-island}}=2\pi \int_{r_0}^{r_{\text{UV}}} \sinh (r) \sqrt{1+\frac{ \cosh ^2(r) Z'(r)^2}{Z(r)^2 \left(1-Z(r)^2\right)}} \ dr,
\end{eqnarray}
which takes exactly the same form as (\ref{sect4.1:island area function}) of sect.4.1. Recall from (\ref{sect4.1:bound of zbdy}) that there is a lower bound of the boundary value of $Z$
\begin{eqnarray}\label{sect4.2:bound of Zbdy} 
Z_{\text{bdy}}=Z(r_{UV})\ge z_c \approx 0.886.
\end{eqnarray}
Combining (\ref{sect4.2:bound of Zbdy}) with (\ref{sect4.2: z to Z}), we derive a lower bound of the boundary value of $z$ at time $t_R$ (\ref{sect4.2:tRtL})
\begin{eqnarray}\label{sect4.2:bound of zbdy} 
z_{\text{bdy}}=z_{\text{max}}Z_{bdy} \ge  0.886 z_{\text{max}},
\end{eqnarray}
which yields
\begin{eqnarray}\label{sect4.2:bound of zmax} 
 z_{\text{max}} \le \frac{ z_{\text{bdy}} }{z_c}= \frac{ z_{\text{bdy}} }{0.886} \le \frac{ 1 }{0.886}  \approx 1.128. 
\end{eqnarray}
Above we have used $z_{\text{bdy}}\le 1$ in the second inequality so that the island and bath both lie outside the horizon. From  (\ref{sect4.2:tRtL},\ref{sect4.2:zmaxtR}), we notice that the upper bound of $z_{\text{max}}\le  z_{\text{bdy}} /z_c $ leads to an upper bound of the maximum time in the no-island phase
\begin{eqnarray}\label{sect4.2:bound of tR} 
t_m= \text{max } (t_R=t_L )= \frac{1}{2} \log \left(\frac{z_{\text{bdy}} \sqrt{\frac{1-z_c^2}{z_{\text{bdy}}^2-z_c^2}}+1}{z_{\text{bdy}} \sqrt{\frac{1-z_c^2}{z_{\text{bdy}}^2-z_c^2}}-1}\right),
\end{eqnarray}
for any given $z_{\text{bdy}}$ obeying $z_c\approx 0.886\le z_{\text{bdy}} <1$. Only in the case $z_{\text{bdy}}=1$, the maximum time $t_m$ can become infinity. See Fig. \ref{tmzbdy}. This is quite different from the case of codim-1 brane in AdS/BCFT.  Now we finish the proof of the statement that the extremal surface passing through the horizon cannot be defined after some finite time $t_m$. 

\begin{figure}[t]
\centering
\includegraphics[width=15cm]{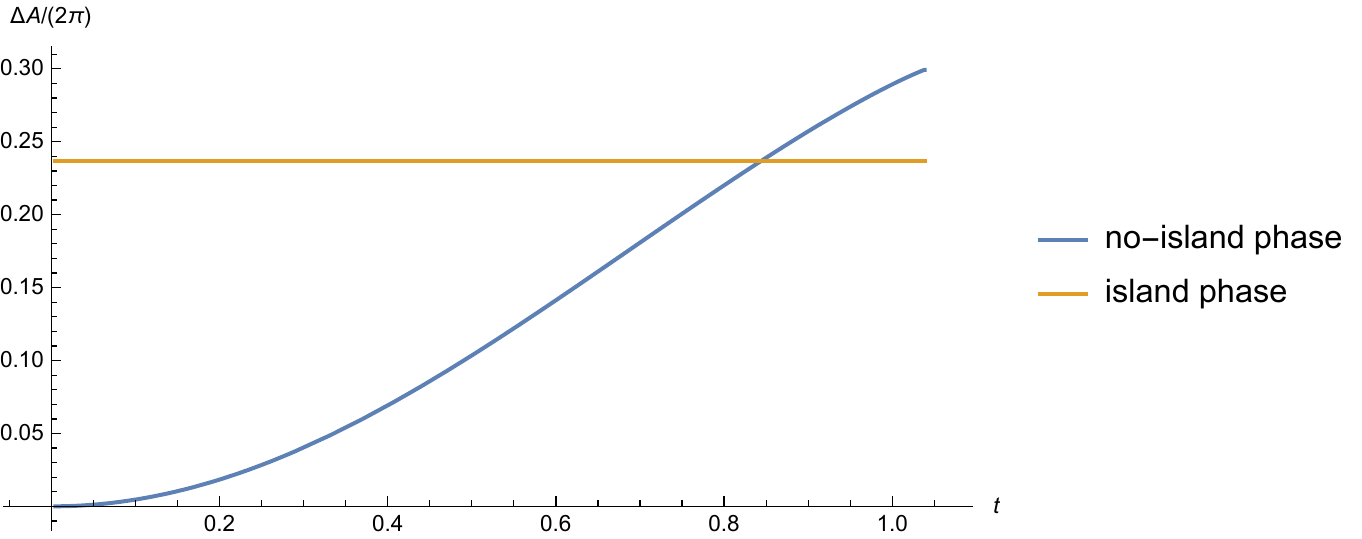}
\caption{Page curve on codim-2 brane in $\text{AdS}_4/\text{dCFT}_3$. 
Here we define $\Delta A=A(t_R)-A(0)$
, choose $z_{\text{bdy}}=0.95 $ and use $t$ to label the boundary time $t_R=t_L$ for simplicity. 
The blue line denotes the area of extremal surface in the no-island phase, which increases with time and is not well-defined after $t_m \approx 1.038$. The yellow line denotes the area of extremal surface in the island phase, which is constant and is always well-defined. 
The holographic entanglement entropy is dominated by the extremal surface with the smaller area. Thus the entanglement entropy first increases with time (blue line) and then becomes a constant (yellow line). In this way, the Page curve of eternal black hole is recovered. Note that the Page time $t_P\approx 0.844$ is smaller than $t_m \approx 1.038$. Thus the unusual situation happens at $t_m$ does not affect the Page curve. }
\label{Pagecurve3d}
\end{figure}

From (\ref{sect4.2:tRtL},\ref{sect4.2: z to Z},\ref{sect4.2: area function Zr}) and the numerical results of $z(r)$ of sect.4.1, we can work out the time evolution of the area of RT surfaces in the no-island phase. See blue line of  Fig. \ref{Pagecurve3d}, which increases with time and 
stops at a finite time $t_m$. Here $\Delta A=A(t_R)-A(0)$ is defined by the difference between the area at time $t_R$ and that at the beginning.  Without loss of generality, we choose 
$z_{\text{bdy}}=0.95 $ which corresponds to the maximum time $t_m \approx 1.038$, after which the extremal surface passing through the horizon is not well-defined.  The yellow line of Fig. \ref{Pagecurve3d} denotes the area of RT surface in the island phase, which is a constant
and is well-defined all the time. 
  At the early times, the blue line has smaller area and thus dominates the entanglement entropy. After the Page time $t_P\approx 0.844$, the yellow line has smaller area and entanglement entropy becomes a constant. The Page curve is given by the blue line for $t\le t_P$ and the yellow line for $t>t_P$.  Note that $t_m \approx 1.038$ is larger than the Page time $t_P\approx 0.844$, thus the unusual situation happens at $t_m$ does not affect the Page curve. 
 It is reminiscent of the principle of cosmic supervision: although there is singularity inside the horizon, the observer outside does not see any non-physical situation.
 Note that there is no singularity in our case.
It should be mentioned that the finiteness of $t_m$ also appears in Gauss Bonnet gravity for a range of couplings obeying the causal constraint for codim-1 brane \cite{Hu:2022ymx}. Unlike the present case, there is a zeroth-order phase transition of entanglement entropy for codim-1 branes in Gauss Bonnet gravity \cite{Hu:2022ymx}.

\subsection{Tensive brane}
In the above section, we focus on the tensionless brane. Let us go on to study the brane with non-zero tension. We find that the larger the tension is, the larger the maximum time of the no-island phase is, and the larger the Page time is. In the large tension limit, the Page time approaches the maximum time of the no-island phase from below. Since the calculations are similar to the case of tensionless branes, we only show the main results below. 

We take the following bulk metric 
\begin{eqnarray}\label{sect4.3:metric for tensive brane}
ds^2=\frac{d\bar{r}^2}{\bar{r}^2-1-\frac{\bar{r}_h(\bar{r}_h^2-1)}{\bar{r}}}+\left(\bar{r}^2-1-\frac{\bar{r}_h(\bar{r}_h^2-1)}{\bar{r}} \right)d\theta^2+ \bar{r}^2 \frac{\frac{dz^2}{1-z^2}-(1-z^2) dt^2}{z^2},
\end{eqnarray}
where the codim-2 brane is located at $\bar{r}=\bar{r}_h$, the AdS boundary is at $\bar{r}\to \infty$ and $\bar{r}_h$ is a constant
\begin{eqnarray}\label{sect4.3:rbar}
\bar{r}_h=\frac{1+\sqrt{1+3 q^2}}{3 q},\ \ q\ge 1,
\end{eqnarray}
which is related to the tension (\ref{tension}) of the brane. Following the approach of sect.4.1 and sect.4.2, we obtain the following results for tensive branes with $q>1$.

{\bf 1 island phase:}

The RT surface ending on the brane is still given by 
\begin{eqnarray}\label{sect4.3:embedding function island}
t=\text{constant}, \ \ z=z_{\text{bdy}}.
\end{eqnarray}
And the corresponding area of RT surface is given by
\begin{eqnarray}\label{sect4.3:area island} 
A_{\text{island}}&=&2\pi q \int_{\bar{r}_h}^{\bar{r}_{\text{UV}}} \sqrt{1+\frac{\bar{r} \left(\bar{r}_h^3-\bar{r}_h-\bar{r}^3+\bar{r}\right) z'\left(\bar{r}\right)^2}{z\left(\bar{r}\right)^2 \left(z\left(\bar{r}\right)^2-1\right)}} d\bar{r}\nonumber\\
&=& 2\pi q (\bar{r}_{\text{UV}}-\bar{r}_h),
\end{eqnarray}
where $\bar{r}_{\text{UV}}$ is a UV cut-off and the period of $\theta$ is $2\pi q$.

{\bf 2 no-island phase at $t_R=t_L=0$: }

\begin{figure}[t]
\centering
\includegraphics[width=7.5cm]{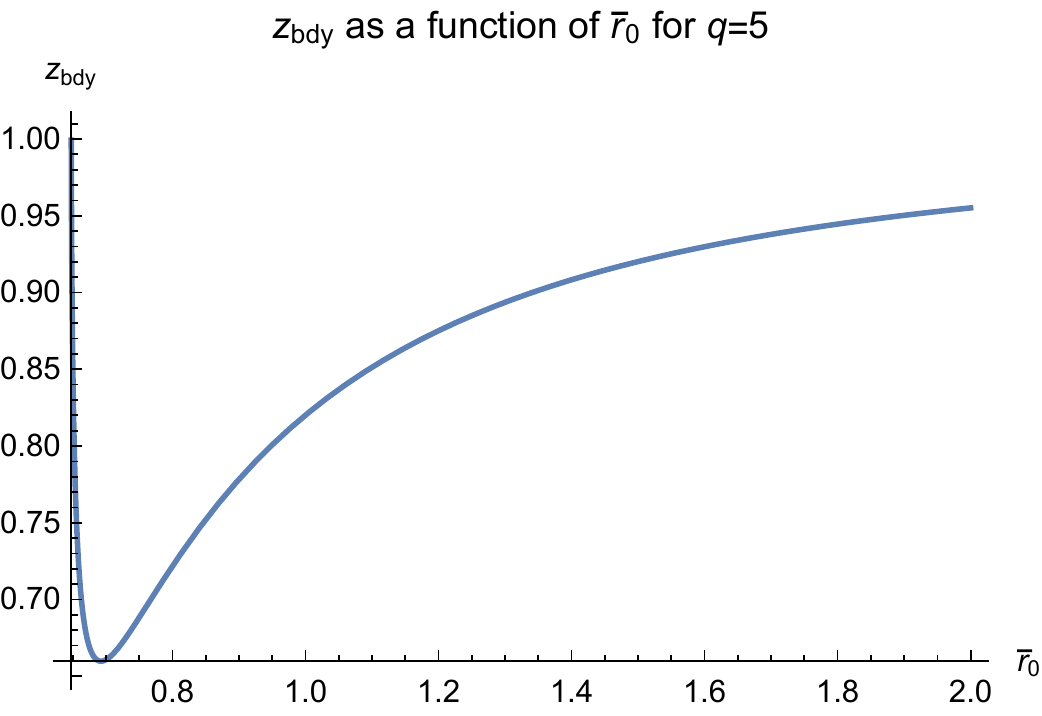} \includegraphics[width=7.5cm]{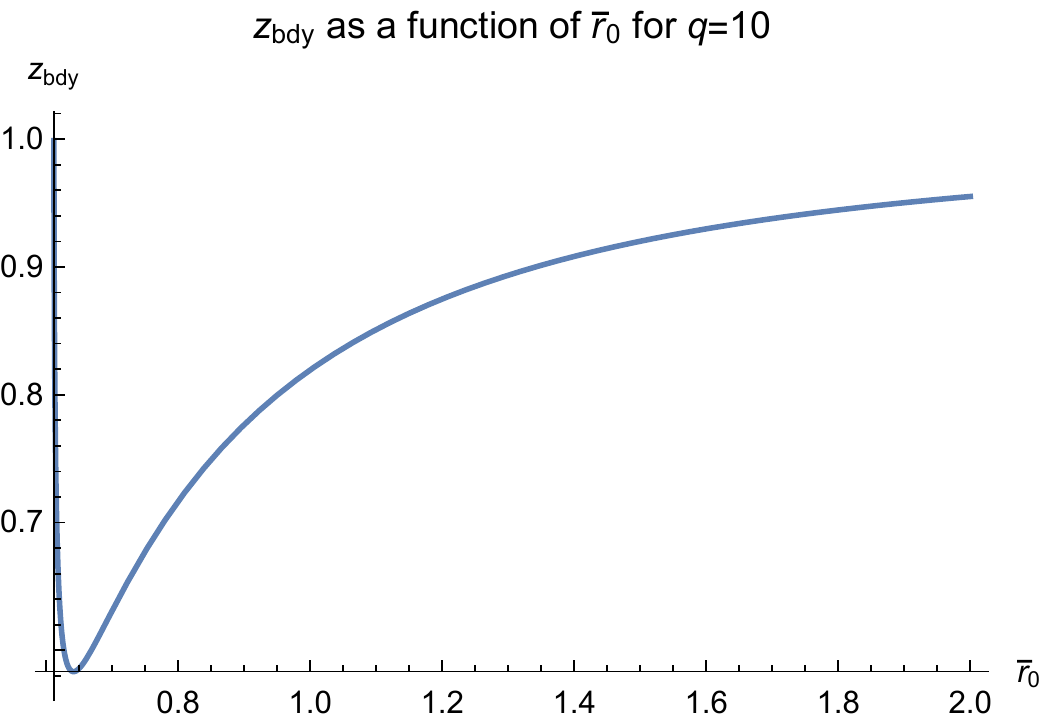}
\caption{$z_{\text{bdy}}$ as a function of $\bar{r}_0$ for $q=5$ (left) and $q=10$ (right) in no-island phase. The lower bound of $z_{\text{bdy}}$ is given by 
$z_c\approx 0.660$ for $q=5$ and $z_c\approx0.583$ for $q=10$, respectively. The larger the tension $q$ is, the smaller the lower bound $z_c$ is. }
\label{zbdyr0q5q10}
\end{figure}

The area functional of RT surface in the no-island phase at the beginning is given by
\begin{eqnarray}\label{sect4.3:area noisland beginning} 
A_{\text{no-island}}=2\pi q \int_{\bar{r}_0}^{\bar{r}_{\text{UV}}} \sqrt{1+\frac{\bar{r} \left(\bar{r}_h^3-\bar{r}_h-\bar{r}^3+\bar{r}\right) z'\left(\bar{r}\right)^2}{z\left(\bar{r}\right)^2 \left(z\left(\bar{r}\right)^2-1\right)}} d\bar{r},
\end{eqnarray}
where $\bar{r}_0\ge \bar{r}_h$ is the value of $\bar{r}$ on the horizon $z=1$.  Taking variations of (\ref{sect4.3:area noisland beginning}), we get EOM
\begin{eqnarray}\label{sect4.3: EOM noisland0} 
z''(\bar{r})&=&\frac{\left(-\bar{r}_h^3+\bar{r}_h+4 \bar{r}^3-2 \bar{r}\right) z'\left(\bar{r}\right)^3}{2 z\left(\bar{r}\right)^2 \left(z\left(\bar{r}\right)^2-1\right)}\nonumber\\
&&+\frac{\left(\bar{r}_h^3-\bar{r}_h-4 \bar{r}^3+2 \bar{r}\right) z'\left(\bar{r}\right)}{\bar{r} \left(-\bar{r}_h^3+\bar{r}_h+\bar{r}^3-\bar{r}\right)}+\frac{\left(2 z\left(\bar{r}\right)^2-1\right) z'\left(\bar{r}\right)^2}{z\left(\bar{r}\right)^3-z\left(\bar{r}\right)}.\end{eqnarray}
Solving the above equation around horizon, we get 
\begin{eqnarray}\label{sect4.3: perturbative solution} 
z(\bar{r})=1+ c_1 (\bar{r}-\bar{r}_0)+ c_2 (\bar{r}-\bar{r}_0)^2+O (\bar{r}-\bar{r}_0)^2,
\end{eqnarray}
where 
\begin{eqnarray}\label{sect4.3: solution c1} 
&&c_1=\frac{2}{\bar{r}_h^3-\bar{r}_h-4 \bar{r}_0^3+2 \bar{r}_0},\\
&&c_2=\frac{-28 \bar{r}_0^3 \bar{r}_h \left(\bar{r}_h^2-1\right)+2 \bar{r}_h^2 \left(\bar{r}_h^2-1\right){}^2+44 \bar{r}_0^6-36 \bar{r}_0^4}{3 \bar{r}_0 \left(-\bar{r}_h^3+\bar{r}_h+\bar{r}_0^3-\bar{r}_0\right) \left(\bar{r}_h^3-\bar{r}_h-4 \bar{r}_0^3+2 \bar{r}_0\right){}^2}. \label{sect4.3: solution c2} 
\end{eqnarray}
From EOM (\ref{sect4.3: EOM noisland0}) and BC (\ref{sect4.3: perturbative solution}), we can solve $z(\bar{r})$ numerically. Similar to the tensionless case, there is a lower bound of $z_{\text{bdy}}$ on the AdS boundary. See Fig. \ref{zbdyr0q5q10} for example, where we have
\begin{eqnarray}\label{sect4.3: bound of zbdy} 
z_{\text{bdy}}\ge z_c\approx\begin{cases}
0. 886, \ \text{for}\  q=1,\\
0. 660, \ \text{for}\  q=5,\\
0.583, \ \text{for}\  q=10,
\end{cases}
\end{eqnarray}
where $q$ labels the tension (\ref{tension}) and $q=1$ corresponds to the tensionless case. It is found that the larger the tension is, the smaller the lower bound of  $z_{\text{bdy}}$ is. 

{\bf 3 no-island phase at $t_R=t_L>0$: }

Consider the embedding functions of the RT surface $ \bar{r}=\bar{r}(z)$ and $v=v(z)$. 
Similar to the tensionless case, the EOM of $v=v(z)$ is decoupled with $r(z)$ and is still given by (\ref{sect4.2:EOMV}) in the tensive case. Thus $v(z)$ (\ref{sect4.2:V}), $t(z)$ (\ref{sect4.2:T}) and $t_R=t_L$ (\ref{sect4.2:tRtL}) derived in sect. 4.1 also apply to the tensive case.  Substituting $v(z)$ into the area functional of RT surface and transforming 
$\bar{r}(z)$ into $z(\bar{r})$, we obtain
\begin{eqnarray}\label{sect4.3: area noisland time} 
A_{\text{no-island}}=2\pi q \int_{\bar{r}_0}^{\bar{r}_{\text{UV}}} \sqrt{1+\frac{z_{\text{max}}^2 \bar{r} \left(-\bar{r}_h^3+\bar{r}_h+\bar{r}^3-\bar{r}\right) z'\left(\bar{r}\right)^2}{z\left(\bar{r}\right)^2 \left(z_{\text{max}}^2-z\left(\bar{r}\right)^2\right)}} \ d\bar{r},
\end{eqnarray}
which agrees with (\ref{sect4.3:area noisland beginning}) at $t_R=t_L=0$ ($z_{\text{max}}=1$).  
Changing $z(\bar{r})\to z_{\text{max}} z(\bar{r})$, the area functional (\ref{sect4.3: area noisland time}) at $t_R=t_L>0$ becomes exactly the same as the area functional (\ref{sect4.3:area noisland beginning}) at $t_R=t_L=0$, which is similar to the tensionless case. Following approach of sect.4.2, we obtain a upper bound of $z_{\text{max}}$
\begin{eqnarray}\label{sect4.3: zmax bound} 
z_{\text{max}} \le \frac{z_{\text{bdy}}}{z_c} \le \frac{1}{z_c} ,
\end{eqnarray}
and a upper bound of the time in the no-island phase
\begin{eqnarray}\label{sect4.3:bound of tR} 
t_m= \text{max } (t_R=t_L )= \frac{1}{2} \log \left(\frac{z_{\text{bdy}} \sqrt{\frac{1-z_c^2}{z_{\text{bdy}}^2-z_c^2}}+1}{z_{\text{bdy}} \sqrt{\frac{1-z_c^2}{z_{\text{bdy}}^2-z_c^2}}-1}\right).
\end{eqnarray}
Since $z_c$ (\ref{sect4.3: bound of zbdy}) becomes smaller in the tensive case, the maximum time $t_m$ becomes larger in the tensive case. Note that there is no bound of time in the island phase. As a result, the entanglement entropy is always well-defined, and our models have no non-physical situation. 

\begin{figure}[t]
\centering
\includegraphics[width=15cm]{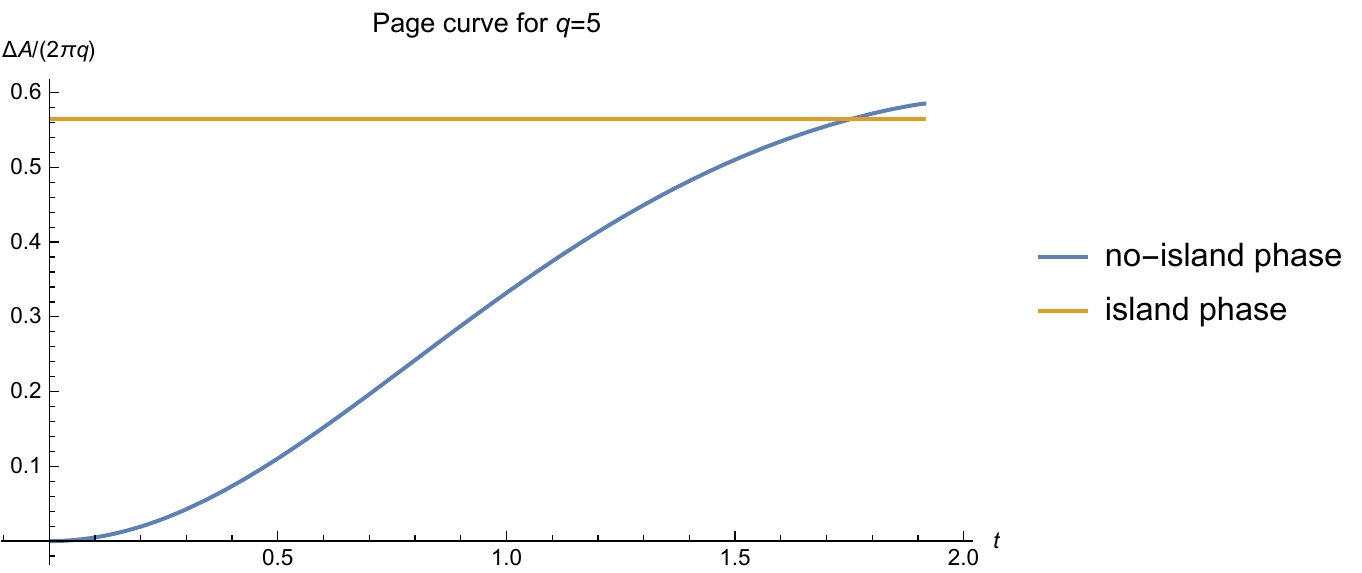}
\caption{ Page curve on tensive codim-2 brane with $q=5$ and
$z_{\text{bdy}}=0.95$ in $\text{AdS}_4/\text{dCFT}_3$. The blue line and yellow line denote the no-island phase and the island phase, respectively. The Page curve is given by the blue line for 
$t\le t_P\approx 1.754$, 
and is given by the yellow line for $t> t_P$. Since the Page time 
$t_P\approx 1.754$ is smaller than the maximum time $t_m \approx 1.914$ in no-island phase, the unusual situation happens at $t_m$ does not affect the Page curve.}
\label{Pagecurve5q}
\end{figure}

\begin{figure}[t]
\centering
\includegraphics[width=15cm]{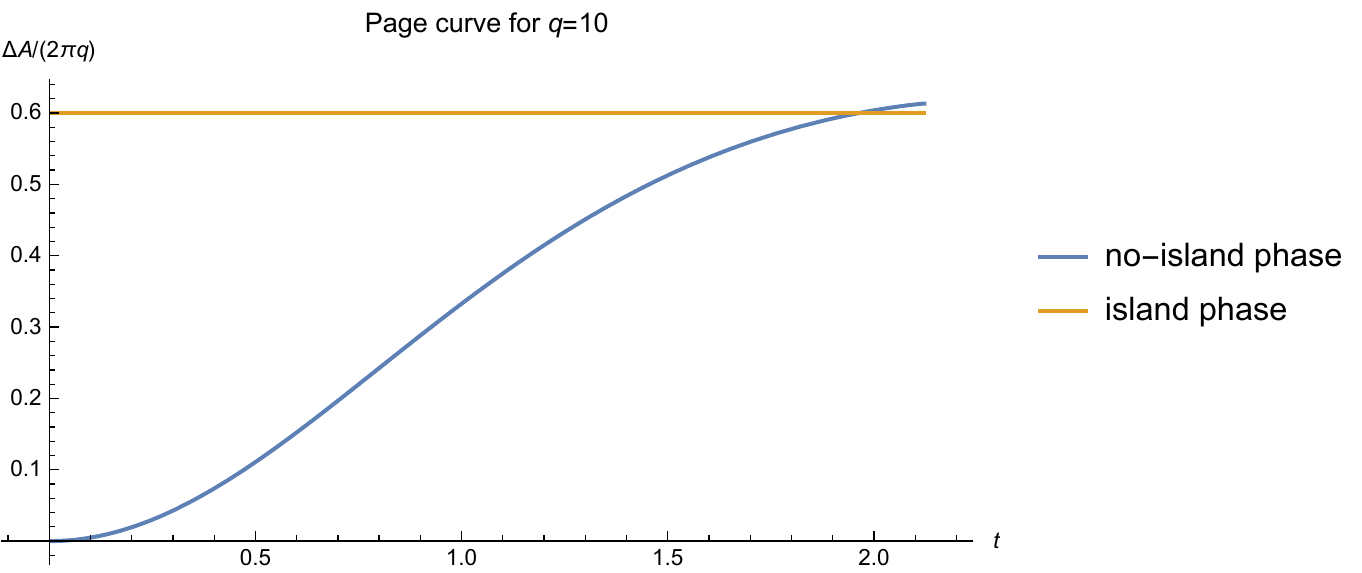}
\caption{ Page curve  tensive codim-2 brane with $q=10$ and
$z_{\text{bdy}}=0.95$ in $\text{AdS}_4/\text{dCFT}_3$. The blue line and yellow line denote the no-island phase and the island phase, respectively. The Page curve is given by the blue line for 
$t\le t_P\approx 1.969$, and is given by the yellow line for $t> t_P$. Since the Page time 
$t_P\approx 1.969$ is smaller than the maximum time $t_m \approx 2.123$ in no-island phase, the unusual situation happens at $t_m$ does not affect the Page curve.}
\label{Pagecurve10q}
\end{figure}

To end this section, we draw the Page curves on the codim-2 branes with non-zero tensions.  See Fig. \ref{Pagecurve5q} and Fig. \ref{Pagecurve10q}. It is found that the larger the tension is, the larger the Page time $t_P$ and the maximum time $t_m $ in the no-island phase is. In the large tension limit, $t_P$ approaches $t_m$ from below. Similar to the tensionless case, since Page time $t_P$ is smaller than the maximum time $t_m $ in no-island phase, the unusual situation happens at $t_m$ does not affect the Page curve. Now we finish the discussions of Page curve of eternal black holes on codim-2 branes with non-zero tensions in $\text{AdS}_4/\text{dCFT}_3$.

\section{Page curve on codim-2 brane in $\text{AdS}_{d+1}/\text{dCFT}_d$}

In this section, we study the island and Page curve on codim-2 branes in higher dimensional AdS$_{d+1}$/dCFT$_{d}$. We consider both hyperbolic black holes and AdS black holes on the brane. Unlike AdS$_4$/dCFT$_3$, it is difficult to obtain analytical results for $d>3$, since the EOMs of $v(z)$ and $r(z)$ are coupled in higher dimensions. Thus, we focus on numeral calculations in this section. We find that the qualitative behavior of Page curve is the same as that of AdS$_4$/dCFT$_3$.
Readers not interested in numerical calculations can skip this chapter.

\subsection{Hyperbolic black hole}
We first study the case of a hyperbolic black hole on a tensionless brane. The corresponding bulk metric is given by
\begin{eqnarray} \label{hyperbolic_metric}
ds^2=   dr^2  +\sinh^2(r) d \theta^2  +\frac{\cosh^2(r)}{z^2} \left(-(1-z^{2})dt^2+\frac{dz^2}{1-z^{2}}+dH_{d-3}^2\right),
\end{eqnarray}
where $dH_{d-3}^2$ is the hyperbolic spatial geometry with unit curvature, i.e.
\begin{equation}
dH_{d-3}^2  = d\chi^2 + \cosh^2 \chi\, dH_{d-4}^2 .
\end{equation}
In the Eddington-Finkelstein coordinates, the bulk metric becomes
\begin{eqnarray} \label{hyperbolic_metric_vz}
ds^2=   dr^2  +\sinh^2(r) d \theta^2  +\frac{\cosh^2(r)}{z^2} \left(-(1-z^{2})dv^2-2dvdz+dH_{d-3}^2\right).
\end{eqnarray}
It is convenient to use the above metric to study the RT surface passing through the horizon in the no-island phase.

\subsubsection{Island phase}
Let us first discuss the island phase, where the RT surface is perpendicular to both the brane and the AdS boundary. This kind of  RT surface is outside the horizon and does not evolve over time.
Substituting the embedding functions (\ref{sect4.1:embedding function}) into the metric (\ref{hyperbolic_metric}), we get the area of RT surface
\begin{eqnarray} \label{hyperbolic_island_Area}
A = 2\pi \text{vol}_{H_{d-3}}\int_{0}^{r_\text{UV}} \frac{ \sinh(r)\cosh^{d-3}(r)}{z(r)^{d-3}} \sqrt{1+\frac{\cosh^2(r)z'(r)^2}{z(r)^2-z(r)^4}}dr.
\end{eqnarray}
where $\text{vol}_{H_{d-3}}=\int dH_{d-3}$ is the volume of horizontal space. Taking variations of (\ref{hyperbolic_island_Area}), we get EOM
\begin{eqnarray} \label{hyperbolic_island_EOM}
z''(r) &=&
\frac{\left((d-5) z(r)^2-d+4\right) z'(r)^2}{z(r) \left(1-z(r)^2\right)}-\frac{\cosh ^2(r) \text{csch}(2 r) ((d-1) \cosh (2 r)-d+3) z'(r)^3}{z(r)^2 \left(1-z(r)^2\right)}\notag\\
&-&\text{csch}(2 r) (d \cosh (2 r)-d+2) z'(r)-2 (d-3) z(r) \left(1-z(r)^2\right) \tanh (r) \text{csch}(2 r).
\end{eqnarray}
Except $z(r)=1$, there is no exact solution to (\ref{hyperbolic_island_EOM}). Thus we have to solve (\ref{hyperbolic_island_EOM}) numerically.  To do so, we need to impose suitable BCs on the brane. We choose the Neumann boundary condition (NBC), which means the RT surface is orthogonal to the brane, i.e, $ z'(0)=0$.  Solving EOM around the brane $r=0$ with NBC $z'(0)=0$, we get
\begin{eqnarray} \label{sect5.1.1: NBC on brane}
z(r)=z_{\text{brane}}+\frac{1}{4} (d-3) z_{\text{brane}} \left(z_{\text{brane}}^2-1\right)r^2+ O(r^4),
\end{eqnarray}
where $z_{\text{brane}}$ is the value of $z$ on the brane, which is related to $z_\text{bdy}$ on the AdS boundary. 
For any given $z_{\text{brane}}$, we can solve EOM (\ref{hyperbolic_island_EOM}) with BC (\ref{sect5.1.1: NBC on brane}) numerically, and then obtain $z(r)$ and $z_\text{bdy}=z(r_{\text{UV}})$.  
For a fixed $z_\text{bdy}$, we can determine the input parameter $z_{\text{brane}}$  by the shooting method. 

Let us go on to study the no-island phase at $t_R=t_L=0$. In this case, the RT surface is perpendicular to both the AdS boundary and the horizon.  Solving (\ref{hyperbolic_island_EOM}) around the horizon, we derive
 \begin{eqnarray} \label{sect5.1.1: NBC on horizon}
z(r)=1-\frac{2 \tanh \left(r_0\right)}{(d-1) \cosh \left(2 r_0\right)-d+3} (r-r_0)+ O(r-r_0)^2,
\end{eqnarray}
which agrees with (\ref{sect4.1:Znearhorizon}) for $d=3$. Recall that $r_0$ is the value of $r$ on horizon, which is related to $z_{\text{bdy}}$. Solving EOM (\ref{hyperbolic_island_EOM}) together with BC (\ref{sect5.1.1: NBC on horizon}) on horizon, we can obtain the RT surface in the no-island phase at $t_R=t_L=0$. Similar to the 3d case, there is a lower bound for $z_{\text{bdy}}$
\begin{eqnarray}\label{sect5.1.1: bound of zbdy} 
z_{\text{bdy}}\ge z_c \approx\begin{cases}
0. 886, \ \text{for}\  d=3,\\
0. 897, \ \text{for}\  d=4,\\
0.902, \ \text{for}\  d=5,
\end{cases}
\end{eqnarray}
which shows that $z_{\text{bdy}}$ increases with the dimensions. Similar to AdS$_4$/dCFT$_3$, the  lower bound of $z_{bdy}$ leads to an upper bound of $z_{\text{max}}$ and the time $t_R=t_L$. We will show this numerically in the following subsection.

\subsubsection{No-island phase}

Let us go on to discuss the no-island phase at $t_R=t_L>0$.  Substituting the embedding functions $r=r(z)$ and $v=v(z)$ into the metric (\ref{hyperbolic_metric_vz}), we derive the area of RT surfaces
\begin{eqnarray} \label{hyperbolic_noisland_Area}
A = 2\pi \text{vol}_{H_{d-3}}\int_{z_\text{bdy}}^{z_{\max}} \frac{ \sinh(r(z))\cosh^{d-3}(r(z))}{z^{d-3}} \sqrt{r'(z)^2-\frac{\cosh^2(r(z))v'(z) \left((1-z^2)v'(z)+2\right)}{z^2} }dz.
\end{eqnarray}
Taking variations of (\ref{hyperbolic_noisland_Area}), we get EOM
\begin{eqnarray} \label{hyperbolic_noisland_EOMv}
v''(z)&=&-\frac{v'(z) \left(\left(d \left(z^2-1\right)-3 z^2+2\right) \left(\left(z^2-1\right) v'(z)-3\right) v'(z)+2 (d-2)\right)}{z} \notag\\
&+&(3-d) z r'(z)^2 \text{sech}^2(r(z)) \left(\left(z^2-1\right) v'(z)-1\right).\\  \label{hyperbolic_noisland_EOMr}
r''(z) &=&-\frac{r'(z) \left(\left(d \left(z^2-1\right)-3 z^2+2\right) \left(\left(z^2-1\right) v'(z)-2\right) v'(z)+2\right)}{z}\notag\\
&-&(d-3) z \left(z^2-1\right) r'(z)^3 \text{sech}^2(r(z))+ r'(z)^2 \text{csch}(2r(z))  (d \cosh (2 r(z))-d+2)\notag\\
&+&\frac{\coth (r(z)) v'(z) \left(\left(z^2-1\right) v'(z)-2\right) ((d-1) \cosh (2 r(z))-d+3)}{2 z^2}.
\end{eqnarray}
Unlike AdS$_4$/dCFT$_3$, $v(z)$ and $r(z)$ are coupled in the above equations, which makes the calculations complicated. 

We impose the following BCs on the turning point $z=z_{\text{max}}$ and $r=r_0$,
\begin{eqnarray}\label{hyperbolic_noisland_BC1}
&&v(z_{\text{max}}) = -\frac{1}{2}\log\frac{z_{\text{max}}+1}{z_{\text{max}}-1} ,   \qquad v'(z_{\max}) = -\infty ,\\
&&r (z_{\max}) = r_{0}, \qquad \qquad \qquad \ \ r'(z_{\max})=\frac{\coth (r_0) ((d-1) \cosh (2 r_0)-(d-3))}{2 z_{\max} \left((d-3) z_{\max}^2-(d-2)\right)}, \label{hyperbolic_noisland_BC2}
\end{eqnarray}
where the first line is just the BC (\ref{sect4.2:zmaxBC}) derived from $t(z_{\text{max}})=0$ and the symmetry of turning point, and the second line can be obtained by solving EOM of $r(z)$.  Let us explain more on how to derive (\ref{hyperbolic_noisland_BC2}).  Since the area functional (\ref{hyperbolic_noisland_Area}) does not depends on $v(z)$ exactly, we can derive a conserved quantity
\begin{eqnarray}\label{sect5.1.2: conserved quantity}
E=\frac{\partial L}{\partial v'(z)}=\frac{z^{1-d} \sinh (r(z)) \left(\left(z^2-1\right) v'(z)-1\right) \cosh ^{d-1}(r(z))}{\sqrt{r'(z)^2+\frac{\cosh ^2(r(z)) v'(z) \left(\left(z^2-1\right) v'(z)-2\right)}{z^2}}},
\end{eqnarray}
where $A = 2\pi \text{vol}_{H_{d-3}} \int_{z_\text{bdy}}^{z_{\max}} L dz$. Substituting  $v'(z_{\max}) = -\infty$ and $r (z_{\max}) = r_{0}$ into the above equation, we derive
\begin{eqnarray}\label{sect5.1.2: conserved quantity1}
E&=&\frac{z^{1-d} \sinh (r(z)) \left(\left(z^2-1\right) v'(z)-1\right) \cosh ^{d-1}(r(z))}{\sqrt{r'(z)^2+\frac{\cosh ^2(r(z)) v'(z) \left(\left(z^2-1\right) v'(z)-2\right)}{z^2}}}\nonumber\\
&=&-\sqrt{z_{\max }^2-1} \sinh \left(r_0\right) \left(z_{\max } \text{sech}\left(r_0\right)\right){}^{2-d}.
\end{eqnarray}
From (\ref{hyperbolic_noisland_EOMr}) and (\ref{sect5.1.2: conserved quantity1}), we can obtain EOM of $r(z)$ which is decoupled with $v(z)$. Solving this decoupled EOM around the turning point, we can derive (\ref{hyperbolic_noisland_BC2}). Note that (\ref{hyperbolic_noisland_BC2}) agrees with (\ref{sect5.1.1: NBC on horizon}) on horizon ($z_{\max }=1$). This can be regarded as a check of our results.  It should be mentioned that, from the decoupled EOM of $r(z)$, we can derive an upper bound of $z_{\max }$ in order to obey the condition $z_{\text{bdy}}\le 1$, which is similar to the case of AdS$_4$/dCFT$_3$. 

We are now ready to solve the RT surfaces in the no-island phase numerically. 
For any given parameters $z_{\max }$ and $r_0$, we can solve EOMs (\ref{hyperbolic_noisland_EOMr}) and (\ref{hyperbolic_noisland_EOMv}) numerically with BCs (\ref{hyperbolic_noisland_BC1}) and (\ref{hyperbolic_noisland_BC2}), and then derive $z_{\text{bdy}}$ from $r(z_{\text{bdy}})=r_{\text{UV}}$ and the boundary time
 \begin{eqnarray}\label{hyperbolic_boundary_time}
t_R = v(z_\text{bdy}) + \frac{1}{2}\log\frac{1+z_{\text{bdy}}}{1-z_{\text{bdy}}} .
\end{eqnarray}
For a fixed $z_{\text{bdy}}$, we have to adjust the input parameters $z_{\max }$ and $r_0$ suitably. It can be achieved by applying the so-called shooting method.  
Note that we should take an UV cut-off for the BC $v'(z_{\max}) = -\infty$ in numerical calculations. This problem can be avoided if we consider the first-order differential equation (\ref{sect5.1.2: conserved quantity1}) instead of the second-order differential equation (\ref{hyperbolic_noisland_EOMv}).

\begin{figure}[!ht]
  \centering
\includegraphics[width=7.5cm]{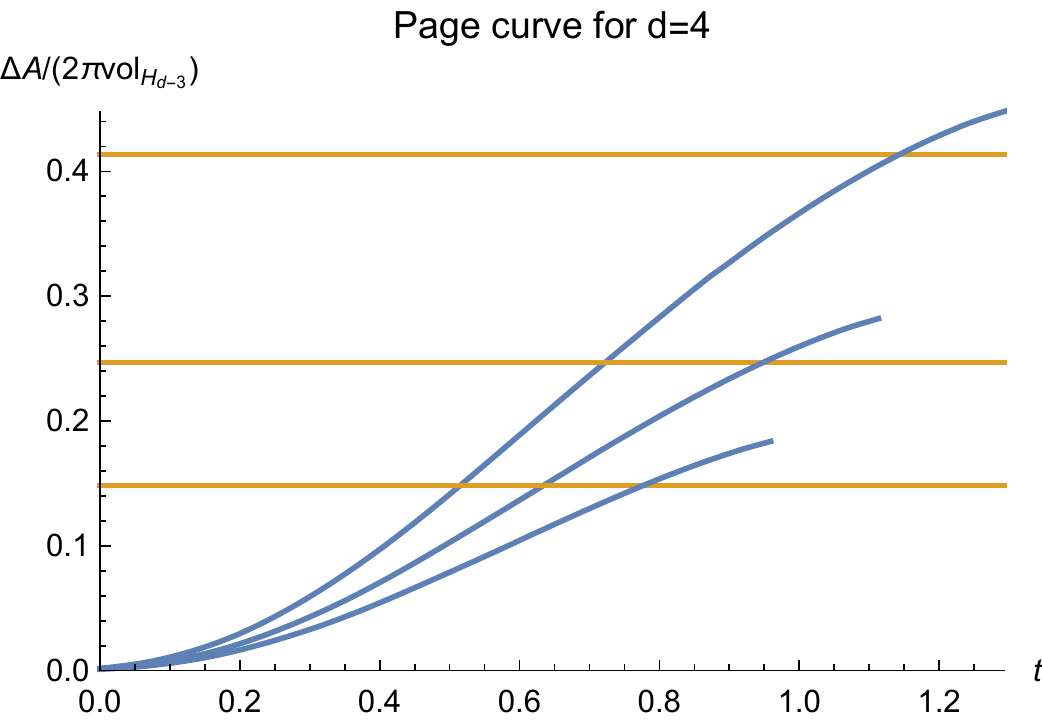}
\includegraphics[width=7.5cm]{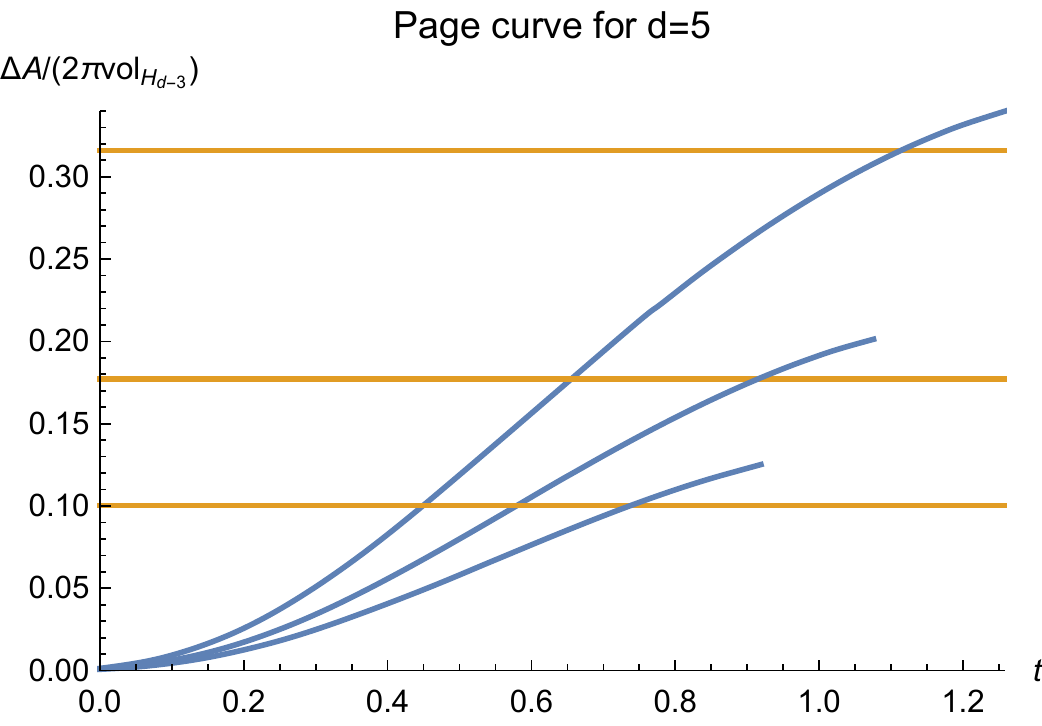} 
\caption{Page curve of hyperbolic black hole on codim-2 tensionless brane in AdS$_5$/dCFT$_4$ (left) and AdS$_6$/dCFT$_5$ (right) for selected 
$z_\text{bdy}=0.95,0.96,0.97$ 
(bottom to top). Here we define $\Delta A=A(t_R)-A(0)$
and label $t_R=t_L$ by $t$ for simplicity. 
The blue line denotes the area of extremal surface in the no-island phase. The yellow line denotes the area of extremal surface in the island phase. The Page curve is given by the blue line for $t\le t_P$ and the yellow line for $t>t_P$. Since we have $t_P< t_m$, the unusual situation happens at $t_m$ does not affect Page curve. }
\label{Pagecurve hyperbolic BH}
\end{figure}

To end this section, let us draw the Page curves for hyperbolic black holes on codim-2 branes with zero tensions. See Fig. \ref{Pagecurve hyperbolic BH}.  Similar to the toy model in AdS$_4$/dCFT$_3$, the Page time $t_P$ is smaller than the maximum time $t_m$ in the no-island phase. As a result, the finite-time problem in the no-island phase does not affect the Page curve.  Following the same approach, we can derive Page curves on the codim-2 brane with non-zero tensions. 
For simplicity, we do not repeat the calculations. See appendix A for some essential formulas, and see Fig. \ref{Pagecurve hyperbolic BH tensive brane} for the results.

\begin{figure}[!ht]
  \centering
\includegraphics[width=7.5cm]{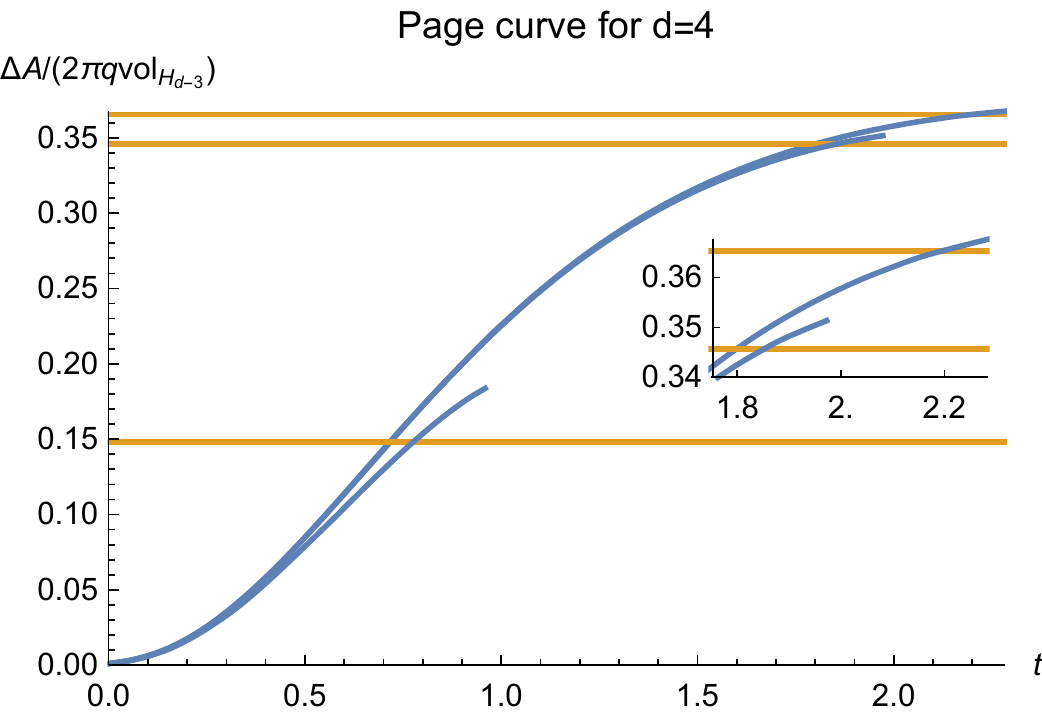}
\includegraphics[width=7.5cm]{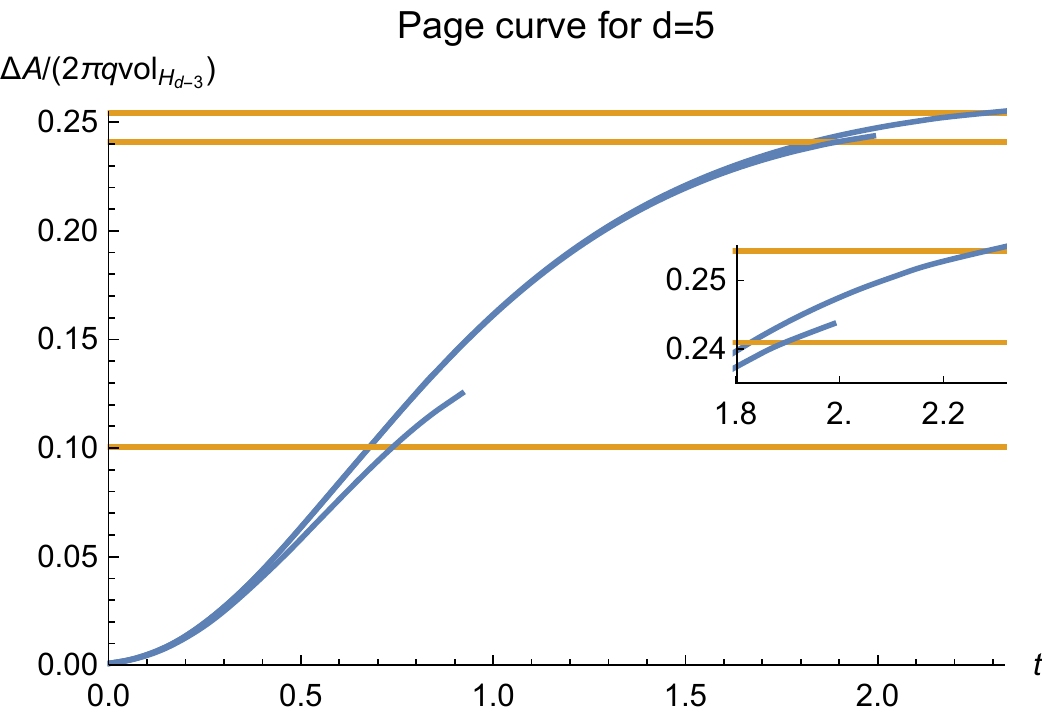} 
\caption{Page curve of hyperbolic black hole on codim-2 tensive brane in AdS$_5$/dCFT$_4$ (left) and AdS$_6$/dCFT$_5$ (right) for selected $q=1,5,10$ (bottom to top) and 
$z_\text{bdy}=0.95$. Here we define $\Delta A=A(t_R)-A(0)$
and label $t_R=t_L$ by $t$ for simplicity. The blue line denotes the area of extremal surface in the no-island phase. The yellow line denotes the area of extremal surface in the island phase. The Page curve is given by the blue line for $t\le t_P$ and the yellow line for $t>t_P$. Since we have $t_P< t_m$, the unusual situation happens at $t_m$ does not affect Page curve. }
\label{Pagecurve hyperbolic BH tensive brane}
\end{figure}

\subsection{AdS black hole}

Now we go on to discuss the case of AdS black hole on a tensionless brane. The bulk metric is given by (\ref{BHmetricr}) with $f(r)=\sinh^2(r)$ and  $g(r)=\cosh^2(r)$. In the Eddington-Finkelstein coordinates, it becomes
\begin{eqnarray}\label{AdSBHmetricvr}
ds^2=dr^2 +\sinh^2(r) d \theta^2  +\frac{\cosh^2(r)}{z^2} \left(-(1-z^{d-2})dv^2-2dvdz+\sum_{a=1}^{d-3}dy_a^2\right).
\end{eqnarray}

\subsubsection{Island Phase}

We discuss the island phase first as before. Substituting the embedding functions (\ref{sect4.1:embedding function}) into the metric (\ref{BHmetricr}) with $f(r)=\sinh^2(r)$ and  $g(r)=\cosh^2(r)$, we get the area of RT surface
\begin{eqnarray} \label{AdS_island_Area}
A = 2\pi \text{vol}_{R_{d-3}}\int_{0}^{r_\text{UV}} \frac{ \sinh(r)\cosh^{d-3}(r)}{z(r)^{d-3}} \sqrt{1+\frac{\cosh^2(r)z'(r)^2}{z(r)^2-z(r)^d}}dr.
\end{eqnarray}
where $\text{vol}_{R_{d-3}}=\int d^{d-3}y$ is the volume of horizontal space.
Taking variations of (\ref{AdS_island_Area}), we get EOM
\begin{eqnarray} \label{AdS_island_EOM}
z''(r) &=&-\frac{(d-6) z'(r)^2}{2 z(r)}-d \coth (2 r) z'(r)+(d-2) \text{csch}(2 r) z'(r)\notag\\
&+&\frac{z'(r)^2 \left(\coth (r) ((d-1) \cosh (2 r)+(3-d)) z'(r)+(d-2) z(r)\right)}{2 z(r)^2 \left(z(r)^{d-2}-1\right)}\notag\\
&+&2 (d-3) \tanh (r) \text{csch}(2 r) \left(z(r)^{d-1}-z(r)\right).
\end{eqnarray}
 Solving EOM around the brane $r=0$ with NBC $z'(0)=0$, we get
\begin{eqnarray} \label{sect5.2.1: NBC on brane}
z(r)=z_{\text{brane}}+\frac{1}{4} (d-3) z_{\text{brane}} \left(z_{\text{brane}}^{d-2}-1\right)r^2+ O(r^4).
\end{eqnarray}
For any given $z_{\text{brane}}$, we can solve EOM (\ref{AdS_island_EOM}) with BC (\ref{sect5.2.1: NBC on brane}) numerically, and then obtain $z(r)$ and $z_\text{bdy}=z(r_{\text{UV}})$. 

Let us go on to study the no-island phase at $t_R=t_L=0$. In this case, the RT surface is perpendicular to both the AdS boundary and the horizon.  Solving (\ref{AdS_island_EOM}) around the horizon, we derive
 \begin{eqnarray} \label{sect5.2.1: NBC on horizon}
z(r)=1-\frac{(d-2) \tanh \left(r_0\right)}{(d-1) \cosh \left(2 r_0\right)-d+3} (r-r_0)+ O(r-r_0)^2.
\end{eqnarray}
Similar to previous cases, there is a lower bound for $z_{\text{bdy}}$
\begin{eqnarray}\label{sect5.2.1: bound of zbdy} 
z_{\text{bdy}}\ge z_c\approx\begin{cases}
0. 897, \ \text{for}\  d=4,\\
0.837, \ \text{for}\  d=5,\\
0.757, \ \text{for}\  d=6.
\end{cases}
\end{eqnarray}
Different from hyperbolic black hole cases, the lower bound of $z_{\text{bdy}}$ (\ref{sect5.2.1: bound of zbdy}) decreases with the dimensions. As in the previous cases, the lower bound of $z_{bdy}$ also leads to an upper bound of $z_{\text{max}}$ and the time $t_R=t_L$.

\subsubsection{No-island Phase}

\begin{figure}[!ht]
  \centering
\includegraphics[width=7.5cm]{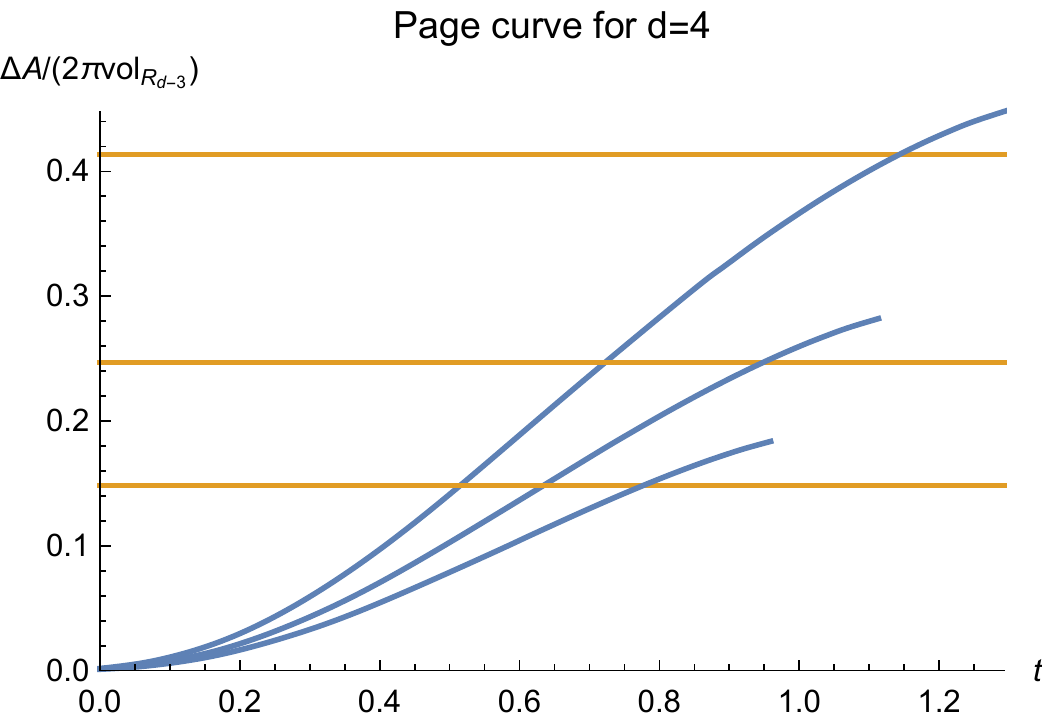}
\includegraphics[width=7.5cm]{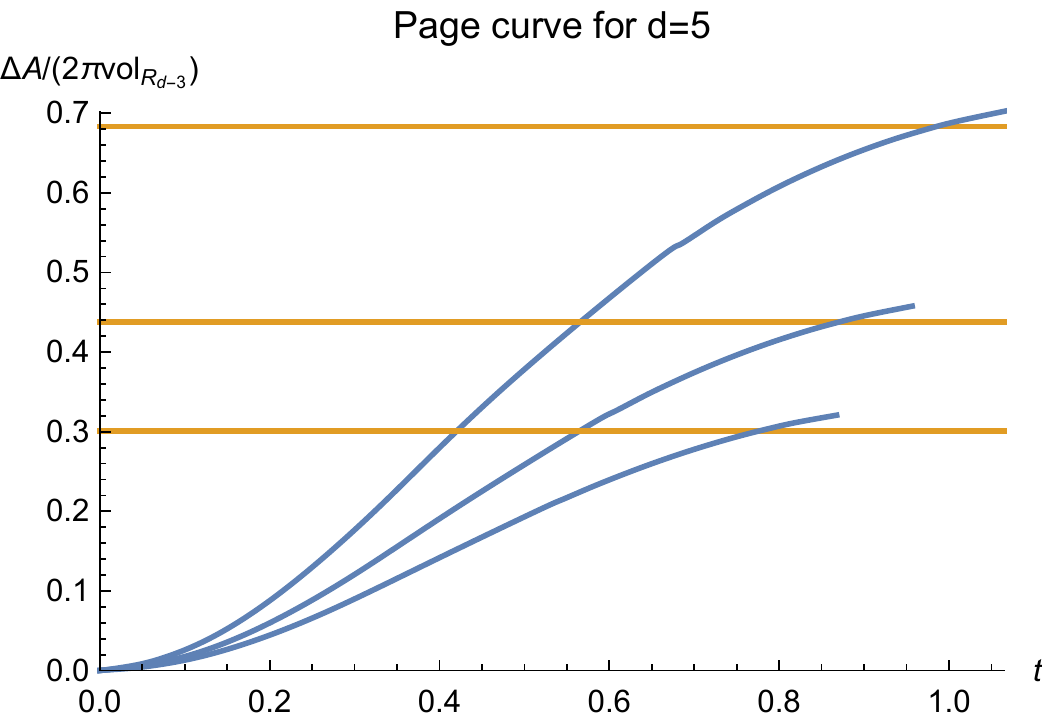}
\caption{Page curve of AdS black hole on codim-2 tensionless brane in AdS$_5$/dCFT$_4$ (left) and AdS$_6$/dCFT$_5$ (right) for selected
$z_\text{bdy}=0.95,0.96,0.97$ (bottom to top). Here we define $\Delta A=A(t_R)-A(0)$. The blue line denotes the area of extremal surface in the no-island phase. The yellow line denotes the area of extremal surface in the island phase. The Page curve is given by the blue line for $t\le t_P$ and the yellow line for $t>t_P$. Since we have $t_P< t_m$, the unusual situation happens at $t_m$ does not affect Page curve. }
\label{Pagecurve AdS BH}
\end{figure}

\begin{figure}[!ht]
  \centering
\includegraphics[width=7.5cm]{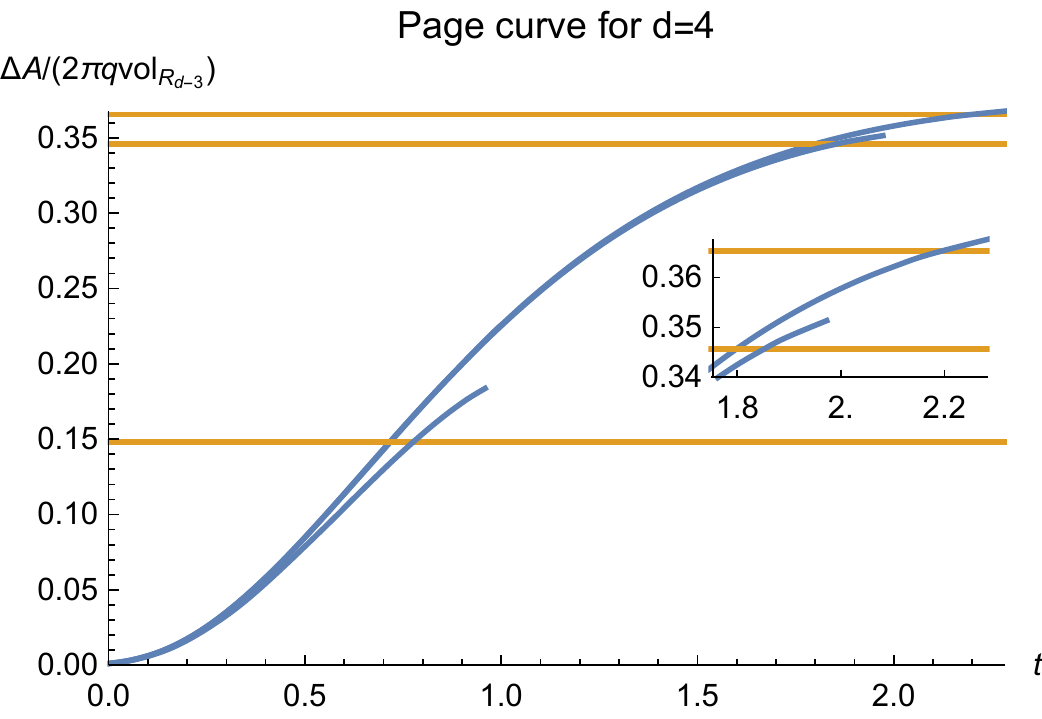}
\includegraphics[width=7.5cm]{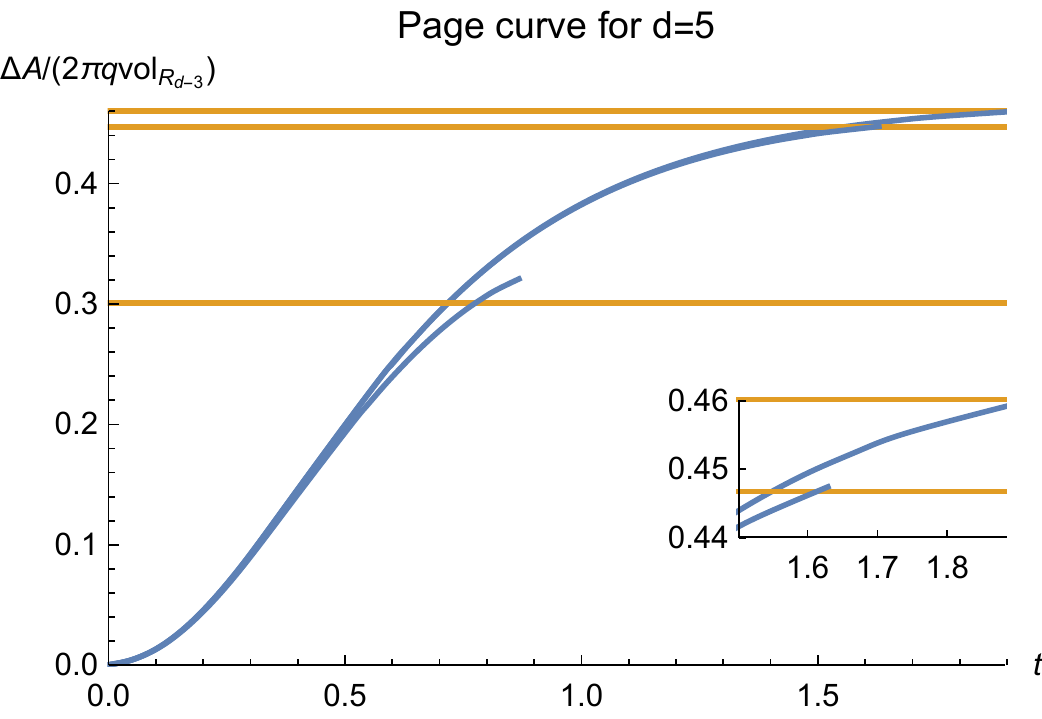} 
\caption{Page curve of AdS black hole on codim-2 tensive brane in AdS$_5$/dCFT$_4$ (left) and AdS$_6$/dCFT$_5$ (right) for selected $q=1,5,10$ (bottom to top) and 
$z_\text{bdy}=0.95$. Here we define $\Delta A=A(t_R)-A(0)$. The blue line denotes the area of extremal surface in the no-island phase. The yellow line denotes the area of extremal surface in the island phase. The Page curve is given by the blue line for $t\le t_P$ and the yellow line for $t>t_P$. Since we have $t_P< t_m$, the unusual situation happens at $t_m$ does not affect Page curve. }
\label{Pagecurve AdS BH tensive brane}
\end{figure}

Let us go on the discuss the no-island phase. Substituting the embedding functions 
 $r=r(z)$ and $v=v(z)$ into the metric (\ref{AdSBHmetricvr}), we derive the area of RT surfaces
\begin{eqnarray} \label{AdS_noisland_Area}
A = 2\pi \text{vol}_{R_{d-3}}\int_{z_\text{bdy}}^{z_{\max}} \frac{ \sinh(r(z))\cosh^{d-3}(r(z))}{z^{d-3}} \sqrt{r'(z)^2-\frac{\cosh^2(r(z))v'(z) \left((1-z^{d-2})v'(z)+2\right)}{z^2} }dz.\nonumber\\
\end{eqnarray}
Taking variations of (\ref{AdS_noisland_Area}), we get 
\begin{eqnarray} 
v''(z)&=&-\frac{(d-2) v'(z) \left(\left(z^{2 d-4}-3 z^{d-2}+2\right) v'(z)^2+\left(6-3 z^{d-2}\right) v'(z)+4\right)}{2 z} \notag\\
&+&(d-3) z r'(z)^2 \text{sech}^2(r(z)) \left(\left(1-z^{d-2}\right) v'(z)+1\right),\label{AdS_noisland_EOMv} \\
r''(z)&=&-\frac{r'(z) \left((d-2) \left(z^{2 d-4}-3 z^{d-2}+2\right) v'(z)^2-2 (d-2) \left(z^{d-2}-2\right) v'(z)+4\right)}{2 z}\notag\\
&-&(d-3) z \left(z^{d-2}-1\right) r'(z)^3 \text{sech}^2(r(z))+ r'(z)^2 \text{csch}(2r(z))  (d \cosh (2 r(z))-d+2)\notag\\
&+&\frac{\coth (r(z)) v'(z) \left(\left(z^{d-2}-1\right) v'(z)-2\right) ((d-1) \cosh (2 r(z))-d+3)}{2 z^2}.\label{AdS_noisland_EOMr}
\end{eqnarray}
The above equations can be solved numerically with BCs
 \begin{eqnarray}
&&v(z_{\text{max}}) = -\int_0^{z_{\max}} \frac{dz}{1-z^{d-2}} ,   \qquad v'(z_{\max}) = -\infty ,\label{AdS_noisland_BCv}\\
&&r (z_{\max}) = r_{0}, \qquad \qquad \qquad    r'(z_{\max})=\frac{\coth (r_0) ((d-1) \cosh (2 r_0)-(d-3))}{ (d-2)z_{\max} \left( z_{\max}^{d-2}-2\right)}.\label{AdS_noisland_BCr}
\end{eqnarray}
Note that we should take an UV cut-off for the BC $v'(z_{\max}) = -\infty$ in numerical calculations. 
Note also that (\ref{AdS_noisland_BCr}) can be derived in the same way as (\ref{hyperbolic_noisland_BC2}). 
For any given parameters $z_{\max }$ and $r_0$, we can numerically derive $z_{\text{bdy}}$ from $r(z_{\text{bdy}})=r_{\text{UV}}$ and obtain the boundary time
 \begin{eqnarray}\label{AdS_boundary_time}
t_R = v(z_\text{bdy}) + \int_0^{z_\text{bdy}} \frac{dz}{1-z^{d-2}}.
\end{eqnarray}
For a fixed $z_{\text{bdy}}$ in our case, we should adjust the parameters $z_{\max }$ and $r_0$ suitably by applying the shooting method. 

Let us draw the Page curves for AdS black holes on codim-2 tensionless branes. See Fig. \ref{Pagecurve AdS BH}.   Following the same approach, we can derive Page curves on the codim-2 brane with non-zero tensions.  See appendix B for some key formulas, and see Fig. \ref{Pagecurve AdS BH tensive brane} for the results. Note that the $d=4$ cases in Fig. \ref{Pagecurve AdS BH} and Fig. \ref{Pagecurve AdS BH tensive brane} are exactly the same as the $d=4$ cases in Fig. \ref{Pagecurve hyperbolic BH} and Fig. \ref{Pagecurve hyperbolic BH tensive brane}, since they have the same bulk metrics.

\section{Conclusions and Discussions}

In this paper, we investigate the mass spectrum of gravitons and the island on codim-2 branes in AdS/dCFT. We find that mass spectrum is positive and the massless mode is forbidden by either the boundary or normalization conditions. We show that the first massive gravitational mode is located on the codim-2 brane; the larger the tension, the smaller the mass, the better the localization. It is similar to the case of codim-1 brane and builds an excellent physical foundation for studying black hole evolution on codim-2 branes.  By studying a toy model in AdS$_4$/dCFT$_3$, hyperbolic black holes and AdS black holes in higher dimensional AdS/dCFT, we find that the Page curve of eternal black holes can be recovered due to the island ending on the codim-2 brane. The new feature is that the extremal surface passing the horizon cannot be defined after some finite time in the no-island phase. Fortunately, this unusual situation does not affect the Page curve since it happens after Page time.  

There are many significant problems to explore. First, it is expected that the effective theory on the codim-2 brane in 
AdS$_4$/dCFT$_3$ is JT gravity. An interesting issue is how to derive JT gravity and explore the corresponding island mechanism on the codim-2 brane. Second, this paper focuses on codim-2 branes in Einstein gravity. It is interesting to generalize the discussions to higher codimensional branes in higher derivative gravity. Third, we focus on eternal black holes in this paper. Studying the Page curve of evolving black holes on codim-$n$ branes is also enjoyable. Fourth, AdS/dCFT can be regarded as a particular limit of cone holography \cite{Miao:2021ual}, where the would-be-AdS-boundary brane is located at a finite place. Recall that there is a massless gravitational mode on the brane in cone holography \cite{Miao:2021ual}. Studying the Page curve of massless gravity in cone holography is interesting. Fifth, a physical explanation of the finite-time phenomenon in the no-island phase is also enjoyable. We hope these problems can be addressed in the future.

\section*{Acknowledgements}

Rong-Xin Miao thank the hospitality of organizers and valuable discussions of participants during the workshops ``The 3rd National Symposium on Field Theory and String Theory" in Beijing. This work is supported by the National Natural Science Foundation of China (Grant No.11905297) and Guangdong Basic and Applied Basic Research Foundation (No.2020A1515010900).

\appendix

\section{Hyperbolic black hole on codim-2 tensive brane}

In this appendix, we list the EOM and BC needed for the numeral calculations of the Page curve of a hyperbolic black hole on the codim-2 brane with non-zero tensions. The corresponding bulk metric takes the form
\begin{eqnarray} 
ds^2&=&   \frac{d\bar{r}^2}{F(\bar{r})} +F(\bar{r}) d\theta^2 +  \frac{\bar{r}^2}{z^2} \left(-(1-z^{2})dt^2+\frac{dz^2}{1-z^{2}}+dH_{d-3}^2\right) \label{hyperbolic_tensive_metric}\\
&=& \frac{d\bar{r}^2}{F(\bar{r})} +F(\bar{r}) d\theta^2 +  \frac{\bar{r}^2}{z^2}\left(-(1-z^{2})dv^2-2dvdz+dH_{d-3}^2\right),\label{hyperbolic_tensive_metric_vz}
\end{eqnarray}
where $F(\bar{r})$ is given by (\ref{rbarF}).

\paragraph{Island phase}
The embedding function of RT surfaces are given by
\begin{eqnarray} \label{hyperbolic_tensive_island}
t = \text{constant},\quad  z=z(\bar{r}).
\end{eqnarray}
Substituting (\ref{hyperbolic_tensive_island}) into (\ref{hyperbolic_tensive_metric}), we derive the area of RT surfaces
\begin{eqnarray} \label{hyperbolic_tensive_island_Area}
A = 2\pi q \text{vol}_{H_{d-3}}\int_{\bar{r}_h}^{\bar{r}_\text{UV}} d\bar{r}\, \frac{\bar{r}^{d-3}}{z(\bar{r})^{d-3}} \sqrt{\frac{F(\bar{r})\bar{r}^2 z'(\bar{r})^2}{z(\bar{r})^2 (1-z(\bar{r})^2)}+1 } .
\end{eqnarray}

Taking variations of (\ref{hyperbolic_tensive_island_Area}), we get EOM
\begin{eqnarray} \label{hyperbolic_tensive_island_EOM}
z''(\bar{r}) &=&-\frac{z'(\bar{r}) \left((d-1) F(\bar{r})+\bar{r} F'(\bar{r})\right)}{ \bar{r} F(\bar{r})}+\frac{ (d-3) \left(z(\bar{r})^2-2\right) z(\bar{r})^3}{ \bar{r}^2 F(\bar{r}) \left(z(\bar{r})^2-1\right)}+\frac{(d-4) z'(\bar{r})^2}{z(\bar{r}) \left(z(\bar{r})^2-1\right)}\notag \\
&-&\frac{z(\bar{r}) \left((d-5) \bar{r}^2 F(\bar{r}) z'(\bar{r})^2-d+3\right)}{ \bar{r}^2 F(\bar{r}) \left(z(\bar{r})^2-1\right)}+\frac{\bar{r} z'(\bar{r})^3 \left(2 (d-2) F(\bar{r})+\bar{r} F'(\bar{r})\right)}{2 z(\bar{r})^2 \left(z(\bar{r})^2-1\right)}.
\end{eqnarray}

Solving EOM (\ref{hyperbolic_tensive_island_EOM}) around the brane $\bar{r}=\bar{r}_h$, we get the BC
\begin{eqnarray} \label{hyperbolic_tensive_island_BC}
z(\bar{r})  =  z_\text{brane} + \frac{(d-3)z_\text{brane}(z_\text{brane}^2-1)}{\bar{r}_h(d \bar{r}_h^2-d+2)} (\bar{r}-\bar{r}_h) + O(\bar{r}-\bar{r}_h)^2.
\end{eqnarray}

\paragraph{No-island phase}
Substituting the embedding functions $\bar{r}=\bar{r}(z)$ and $v=v(z)$ into the metric (\ref{hyperbolic_tensive_metric_vz}), we derive the area of RT surfaces
\begin{eqnarray} \label{hyperbolic_tensive_noisland_Area}
A = 2\pi q \text{vol}_{H_{d-3}}\int_{z_\text{bdy}}^{z_{\max}} dz\, \frac{\bar{r}(z)^{d-3}}{z^{d-3}} \sqrt{\frac{-F(\bar{r}(z))\bar{r}(z)^2 v'(z)\left((1-z^2)v'(z)+2\right)}{z^2 }+\bar{r}'(z)^2 }.
\end{eqnarray}

Taking variations of (\ref{hyperbolic_tensive_noisland_Area}), we get EOM
\begin{eqnarray} 
v''(z) &=&-\frac{v'(z) \left(\left(d \left(z^2-1\right)-3 z^2+2\right) \left(\left(z^2-1\right) v'(z)-3\right) v'(z)+2 (d-2)\right)}{z}\notag \\
&-&\frac{(d-3) z \bar{r}'(z)^2 \left(\left(z^2-1\right) v'(z)-1\right)}{\bar{r}(z)^2 F(\bar{r}(z))},\label{hyperbolic_tensive_noisland_EOMv}\\
\bar{r}''(z) &=&-\frac{\bar{r}'(z) \left(\left(d \left(z^2-1\right)-3 z^2+2\right) \left(\left(z^2-1\right) v'(z)-2\right) v'(z)+2\right)}{z}\notag\\
&+&\frac{(d-2) \bar{r}(z) F(\bar{r}(z)) v'(z) \left(\left(z^2-1\right) v'(z)-2\right)}{z^2}+\frac{\bar{r}(z)^2 v'(z) \left(\left(z^2-1\right) v'(z)-2\right) F'(\bar{r}(z))}{2 z^2}\notag\\
&+&\frac{\bar{r}'(z)^2 \left(\bar{r}(z)^2 F'(\bar{r}(z))-(d-3) z \left(z^2-1\right) \bar{r}'(z)\right)}{\bar{r}(z)^2 F(\bar{r}(z))} +\frac{(d-1) \bar{r}'(z)^2}{\bar{r}(z)}.\label{hyperbolic_tensive_noisland_EOMr}
\end{eqnarray}

We can solve (\ref{hyperbolic_tensive_noisland_EOMv}) and (\ref{hyperbolic_tensive_noisland_EOMr}) numerically with the following BCs
 \begin{eqnarray}
&&v(z_{\text{max}}) = -\frac{1}{2}\log\frac{z_{\text{max}}+1}{z_{\text{max}}-1} ,   \qquad v'(z_{\max}) = -\infty ,\label{hyperbolic_tensive_noisland_BCv}\\
&&\bar{r} (z_{\max}) = \bar{r}_{0}, \qquad \qquad \qquad \qquad   \bar{r}'(z_{\max})=\frac{\bar{r}_0 (2(d-2)F(\bar{r}_0)+\bar{r}_0 F'(\bar{r}_0))}{2 z_{\max}((d-3)z_{\max}^{2}-(d-2))}.\label{hyperbolic_tensive_noisland_BCr}
\end{eqnarray}
Note that we should take an UV cut-off for the BC $v'(z_{\max}) = -\infty$ in numerical calculations.

\section{AdS black hole on codim-2 tensive brane}
In this appendix, we list the EOM and BC needed for the numeral calculations of the Page curve of an AdS black hole on the codim-2 brane with non-zero tensions. The bulk metric with an AdS black hole on the brane takes the form
\begin{eqnarray} 
ds^2&=&   \frac{d\bar{r}^2}{F(\bar{r})} +F(\bar{r}) d\theta^2 +  \frac{\bar{r}^2}{z^2} \left(-(1-z^{d-2})dt^2+\frac{dz^2}{1-z^{d-2}}+\sum_{a=1}^{d-3}dy_a^2\right) \label{AdS_tensive_metric}\\
&=& \frac{d\bar{r}^2}{F(\bar{r})} +F(\bar{r}) d\theta^2 +  \frac{\bar{r}^2}{z^2}\left(-(1-z^{d-2})dv^2-2dvdz+\sum_{a=1}^{d-3}dy_a^2\right),\label{AdS_tensive_metric_vz}
\end{eqnarray}
where $F(\bar{r})$ is given by (\ref{rbarF}).

\paragraph{Island phase}
The embedding function of RT surfaces are the same as (\ref{hyperbolic_tensive_island}). Substituting (\ref{hyperbolic_tensive_island}) into (\ref{AdS_tensive_metric}), we derive the area of RT surfaces
\begin{eqnarray} \label{AdS_tensive_island_Area}
A = 2\pi q\text{vol}_{R_{d-3}}\int_{\bar{r}_h}^{\bar{r}_\text{UV}} d\bar{r}\, \frac{\bar{r}^{d-3}}{z(\bar{r})^{d-3}} \sqrt{\frac{F(\bar{r})\bar{r}^2 z'(\bar{r})^2}{z(\bar{r})^2 (1-z(\bar{r})^{d-2})}+1 } .
\end{eqnarray}

Taking variations of (\ref{AdS_tensive_island_Area}), we get EOM
\begin{eqnarray} \label{AdS_tensive_island_EOM}
z''(\bar{r})&=&\frac{z'(\bar{r})^2 \left(\bar{r} z'(\bar{r}) \left(2 (d-2) F(\bar{r})+\bar{r} F'(\bar{r})\right)+(d-2) z(\bar{r})\right)}{2 \left(z(\bar{r})^d-z(\bar{r})^2\right)}+\frac{(d-3) z(\bar{r})^{d-1}}{\bar{r}^2 F(\bar{r})}\notag\\
&-&\frac{(d-1) z'(\bar{r})}{\bar{r}}-\frac{(d-6) z'(\bar{r})^2}{2 z(\bar{r})}-\frac{(d-3) z(\bar{r})+\bar{r}^2 F'(\bar{r}) z'(\bar{r})}{\bar{r}^2 F(\bar{r})}.
\end{eqnarray}

Solving EOM (\ref{AdS_tensive_island_EOM}) around the brane $\bar{r}=\bar{r}_h$, we get the BC
\begin{eqnarray} \label{AdS_tensive_island_BC}
z(\bar{r}_h)  =  z_\text{brane}+ \frac{(d-3)z_\text{brane}(z_\text{brane}^{d-2}-1)}{\bar{r}_h(d \bar{r}_h^2-d+2)} (\bar{r}-\bar{r}_h) + O(\bar{r}-\bar{r}_h)^2.
\end{eqnarray}

\paragraph{No-island phase}
Substituting the embedding functions $\bar{r}=\bar{r}(z)$ and $v=v(z)$ into the metric (\ref{AdS_tensive_metric_vz}), we derive the area of RT surfaces
\begin{eqnarray} \label{AdS_tensive_noisland_Area}
A = 2\pi q\text{vol}_{R_{d-3}}\int_{z_\text{bdy}}^{z_{\max}} dz\, \frac{\bar{r}(z)^{d-3}}{z^{d-3}} \sqrt{\frac{-F(\bar{r}(z))\bar{r}(z)^2 v'(z)\left((1-z^{d-2})v'(z)+2\right)}{z^2 }+\bar{r}'(z)^2 }.
\end{eqnarray}

Taking variations of (\ref{AdS_tensive_island_Area}), we get EOM
\begin{eqnarray} 
v''(z) &=&-\frac{(d-2) v'(z) \left(\left(z^{2 d-4}-3 z^{d-2}+2\right) v'(z)^2+\left(6-3 z^{d-2}\right) v'(z)+4\right)}{2 z}\notag\\
&+&\frac{(d-3) z \bar{r}'(z)^2 \left(\left(1-z^{d-2}\right) v'(z)+1\right)}{\bar{r}(z)^2 F(\bar{r}(z))},\label{AdS_tensive_noisland_EOMv}\\
\bar{r}''(z) &=& -\frac{\bar{r}(z) v'(z) \left(\left(1-z^{d-2}\right) v'(z)+2\right) \left(2 (d-2) F(\bar{r}(z))+\bar{r}(z) F'(\bar{r}(z))\right)}{2 z^2}\notag\\
&-&\frac{\bar{r}'(z) \left((d-2) \left(z^{2 d-4}-3 z^{d-2}+2\right) v'(z)^2-2 (d-2) \left(z^{d-2}-2\right) v'(z)+4\right)}{2 z}\notag\\
&+&\frac{\bar{r}'(z)^2 \left(\bar{r}(z)^2 F'(\bar{r}(z))-(d-3) \left(z^{d-1}-z\right) \bar{r}'(z)\right)}{\bar{r}(z)^2 F(\bar{r}(z))}+\frac{(d-1) \bar{r}'(z)^2}{\bar{r}(z)}.\label{AdS_tensive_noisland_EOMr}
\end{eqnarray}

We can solve (\ref{AdS_tensive_noisland_EOMv}) and (\ref{AdS_tensive_noisland_EOMr}) numerically with following BCs
\begin{eqnarray}
&&v(z_{\text{max}}) = -\int_0^{z_{\max}} \frac{dz}{1-z^{d-2}} ,   \qquad v'(z_{\max}) = -\infty ,\label{AdS_tensive_noisland_BCv}\\
&&\bar{r} (z_{\max}) = \bar{r}_{0}, \qquad \qquad \qquad \qquad  \quad  \bar{r}'(z_{\max})= \frac{\bar{r}_0 (2(d-2)F(\bar{r}_0)+\bar{r}_0 F'(\bar{r}_0))}{(d-2)z_{\max}(z_{\max}^{d-2}-2)}  .\label{AdS_tensive_noisland_BCr}
\end{eqnarray}
Note that we should take an UV cut-off for the BC $v'(z_{\max}) = -\infty$ in numerical calculations.

\end{document}